\documentclass[floatfix,aps,prd,preprint,superscriptaddress,tightenlines,nofootinbib,showpacs]{revtex4}

\usepackage{graphicx}
\usepackage{dcolumn}
\usepackage{bm}
\usepackage{xspace}

\begin{document}

\newcommand{\etal}{{\em et al.}}
\newcommand{\ie}{{\em i.e.}}

\newcommand{\CDF}{{\sc cdf}\xspace}
\newcommand{\FNAL}{{\sc fnal}\xspace}
\newcommand{\CLEO}{{\sc cleo}\xspace}
\newcommand{\CLEOIII}{{\sc cleo iii}\xspace}
\newcommand{\SLAC}{{\sc slac}\xspace}
\newcommand{\LEP}{{\sc lep}\xspace}
\newcommand{\BABAR}{$\cal B${\em a}$\overline{\cal B}${\em ar}\xspace}
\newcommand{\DZero}{\ensuremath{{\rm D}\emptyset}\xspace}
\newcommand{\CERN}{{\sc cern}\xspace}
\newcommand{\Belle}{{\sc Belle}\xspace}
\newcommand{\UAone}{{\sc ua1}\xspace}
\newcommand{\OPAL}{{\sc Opal}\xspace}
\newcommand{\HeraB}{{\sc Hera}-$B$\xspace}

\newcommand{\MINUIT}{{\sc minuit}\xspace}

\newcommand{\BAnd}{{\sc .and.}\xspace}
\newcommand{\BOr}{{\sc .or.}\xspace}

\newcommand{\trkman}{{\sc Trkman}\xspace}
\newcommand{\HBStandardsProd}{{\sc HBStandardsProd}\xspace}
\newcommand{\muOneOK}{{\sc mu1ok}\xspace}
\newcommand{\muTwoOK}{{\sc mu2ok}\xspace}
\newcommand{\eleOK}{{\sc eleok}\xspace}

\newcommand{\clg}[1]{\ensuremath{{\cal #1}}\xspace}

\newcommand{\QCD}{{\sc qcd}\xspace}

\newcommand{\idest}{{\em i.e.,}\xspace}
\newcommand{\eg}{{\em e.g.,}\xspace}

\newcommand{\LambdaQCD}{\ensuremath{\Lambda_{\rm QCD}}\xspace}

\newcommand{\KSht}{\ensuremath{K_{\rm S}}\xspace}

\newcommand{\eeV}{\ensuremath{{\rm eV}}\xspace}

\newcommand{\ekeV}{\ensuremath{{\rm keV}}\xspace}
\newcommand{\pkeV}{\ensuremath{{\rm keV}/c}\xspace}
\newcommand{\mkeV}{\ensuremath{{\rm keV}/c^2}\xspace}

\newcommand{\eMeV}{\ensuremath{{\rm MeV}}\xspace}
\newcommand{\pMeV}{\ensuremath{{\rm MeV}/c}\xspace}
\newcommand{\mMeV}{\ensuremath{{\rm MeV}/c^2}\xspace}

\newcommand{\eGeV}{\ensuremath{{\rm GeV}}\xspace}
\newcommand{\pGeV}{\ensuremath{{\rm GeV}/c}\xspace}
\newcommand{\mGeV}{\ensuremath{{\rm GeV}/c^2}\xspace}

\newcommand{\eTeV}{\ensuremath{{\rm TeV}}\xspace}
\newcommand{\pTeV}{\ensuremath{{\rm TeV}/c}\xspace}
\newcommand{\mTeV}{\ensuremath{{\rm TeV}/c^2}\xspace}

\newcommand{\mbarn}{\ensuremath{{\rm mb}}\xspace}
\newcommand{\mubarn}{\ensuremath{{\rm \mu b}}\xspace}

\newcommand{\tsec}{\ensuremath{{\rm s}}\xspace}

\newcommand{\tpsec}{\ensuremath{{\rm ps}}\xspace}
\newcommand{\xpsec}{\ensuremath{{\rm ps}/c^2}\xspace}
\newcommand{\tnsec}{\ensuremath{{\rm ns}}\xspace}
\newcommand{\xnsec}{\ensuremath{{\rm ns}/c^2}\xspace}
\newcommand{\tmusec}{\ensuremath{{\rm \mu s}}\xspace}
\newcommand{\tmsec}{\ensuremath{{\rm ms}}\xspace}

\newcommand{\MHz}{\ensuremath{{\rm MHz}}\xspace}
\newcommand{\kHz}{\ensuremath{{\rm kHz}}\xspace}
\newcommand{\Hz}{\ensuremath{{\rm Hz}}\xspace}

\newcommand{\fm}{\ensuremath{{\rm fm}}\xspace}
\newcommand{\nm}{\ensuremath{{\rm nm}}\xspace}
\newcommand{\micron}{\ensuremath{{\rm \mu m}}\xspace}
\newcommand{\mm}{\ensuremath{{\rm mm}}\xspace}
\newcommand{\cm}{\ensuremath{{\rm cm}}\xspace}
\newcommand{\m}{\ensuremath{{\rm m}}\xspace}
\newcommand{\Tesla}{\ensuremath{{\rm T}}\xspace}
\newcommand{\degC}{\ensuremath{^\circ\,{\rm C}}\xspace}

\newcommand{\invpb}{\ensuremath{{\rm pb}^{-1}}\xspace}
\newcommand{\invfb}{\ensuremath{{\rm fb}^{-1}}\xspace}

\newcommand{\pT}{\ensuremath{p_{\rm T}}\xspace}

\newcommand{\etaDec}{\ensuremath{\eta^{\prime} \rightarrow \eta \pi \pi}\xspace}
\newcommand{\psiDec}{\ensuremath{\psi^{\prime} \rightarrow J/\psi \pi \pi}\xspace}
\newcommand{\ullDec}{\ensuremath{\Upsilon \rightarrow \ell^+ \ell^-}\xspace}
\newcommand{\uuuDec}{\ensuremath{\Upsilon \rightarrow \mu^+ \mu^-}\xspace}
\newcommand{\ueeDec}{\ensuremath{\Upsilon \rightarrow e^+ e^-}\xspace}
\newcommand{\upsDec}[2]{\ensuremath{\Upsilon({#1}S) \rightarrow \Upsilon({#2}S) \pi \pi}\xspace}
\newcommand{\uuXDec}[2]{\ensuremath{\Upsilon({#1}S) \rightarrow \Upsilon({#2}S) X}\xspace}
\newcommand{\upsXDec}[2]{\ensuremath{\Upsilon({#1}S) \rightarrow \Upsilon({#2}S) X}\xspace}
\newcommand{\upsCDec}[2]{\ensuremath{\Upsilon({#1}S) \rightarrow \Upsilon({#2}S) \pi^+ \pi^-}\xspace}
\newcommand{\upsNDec}[2]{\ensuremath{\Upsilon({#1}S) \rightarrow \Upsilon({#2}S) \pi^0 \pi^0}\xspace}
\newcommand{\upsCDecuu}[2]{\ensuremath{\Upsilon({#1}S) \rightarrow \Upsilon({#2}S) \pi^+ \pi^-; \Upsilon({#2}) \rightarrow \mu^+ \mu^-}\xspace}
\newcommand{\upsNDecuu}[2]{\ensuremath{\Upsilon({#1}S) \rightarrow \Upsilon({#2}S) \pi^0 \pi^0; \Upsilon({#2}) \rightarrow \mu^+ \mu^-}\xspace}
\newcommand{\upsCDecee}[2]{\ensuremath{\Upsilon({#1}S) \rightarrow \Upsilon({#2}S) \pi^+ \pi^-; \Upsilon({#2}) \rightarrow e^+ e^-}\xspace}
\newcommand{\upsNDecee}[2]{\ensuremath{\Upsilon({#1}S) \rightarrow \Upsilon({#2}S) \pi^0 \pi^0; \Upsilon({#2}) \rightarrow e^+ e^-}\xspace}
\newcommand{\upsTrn}[2]{\ensuremath{\Upsilon({#1}S) \rightarrow \Upsilon({#2}S)}\xspace}
\newcommand{\ups}[1]{\ensuremath{\Upsilon({#1}S)}\xspace}

\newcommand{\cthx}{\ensuremath{\cos \theta_X}\xspace}

\newcommand{\ReBA}{\ensuremath{\Re(\clg{B}/\clg{A})}\xspace}
\newcommand{\ImBA}{\ensuremath{\Im(\clg{B}/\clg{A})}\xspace}
\newcommand{\ReCA}{\ensuremath{\Re(\clg{C}/\clg{A})}\xspace}
\newcommand{\ImCA}{\ensuremath{\Im(\clg{C}/\clg{A})}\xspace}

\newcommand{\redMark}{\mbox{\textcolor{red}{$\diamondsuit$}}\xspace}
\newcommand{\arrow}{\ensuremath{\rightarrow}\xspace}
\newcommand{\redArrow}{\mbox{\textcolor{red}{$\rightarrow$}}\xspace}
\newcommand{\redPlus}{\mbox{\bf\textcolor{red}{$+$}}\xspace}
\newcommand{\vdp}[2]{\ensuremath{{#1}\cdot{#2}}\xspace}
\newcommand{\qp}[2]{\ensuremath{\vec{q}_{{#1}\perp{#2}}}\xspace}
\newcommand{\Mvariable}[1]{\ensuremath{{#1}}\xspace}
\newcommand{\bbBar}{\ensuremath{b\bar{b}}\xspace}
\newcommand{\diPi}{\ensuremath{\pi\pi}\xspace}
\newcommand{\diPiC}{\ensuremath{\pi^+\pi^-}\xspace}
\newcommand{\diPiN}{\ensuremath{\pi^0\pi^0}\xspace}

\newcommand{\BoAf}{\ensuremath{\clg{B}/\clg{A}}\xspace}
\newcommand{\BoAt}{\ensuremath{\frac{\clg{B}}{\clg{A}}}\xspace}
\newcommand{\CoAf}{\ensuremath{\clg{C}/\clg{A}}\xspace}
\newcommand{\CoAt}{\ensuremath{\frac{\clg{C}}{\clg{A}}}\xspace}

\newcommand{\nUnits}[2]{\ensuremath{#1\,{\rm #2}}\xspace}
\newcommand{\nError}[2]{\ensuremath{#1\pm #2}\xspace}
\newcommand{\nUnErr}[3]{\ensuremath{(#1\pm #2)\,{\rm #3}}\xspace}

\newcommand{\fa}{{\ensuremath{f_{\clg{A}}}\xspace}}
\newcommand{\fb}{{\ensuremath{f_{\clg{B}}}\xspace}}
\newcommand{\fc}{{\ensuremath{f_{\clg{C}}}\xspace}}

\newcommand{\faa}{{\ensuremath{f_{\clg{AA}}}\xspace}}
\newcommand{\fab}{{\ensuremath{f_{\clg{AB}}}\xspace}}
\newcommand{\fbb}{{\ensuremath{f_{\clg{BB}}}\xspace}}
\newcommand{\fac}{{\ensuremath{f_{\clg{AC}}}\xspace}}
\newcommand{\fbc}{{\ensuremath{f_{\clg{BC}}}\xspace}}
\newcommand{\fcc}{{\ensuremath{f_{\clg{CC}}}\xspace}}

\newcommand{\mUpsPi}{{\ensuremath{M_{\Upsilon\pi}}}}
\newcommand{\mPiPi}{{\ensuremath{M_{\pi\pi}}}}

\newcommand{\mUps}{{\ensuremath{M_{\Upsilon}}}}
\newcommand{\mUpsp}{{\ensuremath{M_{\Upsilon'}}}}
\newcommand{\mPi}{{\ensuremath{M_{\pi}}}}
\newcommand{\mell}{{\ensuremath{M_{\ell}}}}

\newcommand{\epp}{{\ensuremath{\epsilon'}}}
\newcommand{\epu}{{\ensuremath{\epsilon}}}
\newcommand{\pPp}{{\ensuremath{P'}}}
\newcommand{\pPu}{{\ensuremath{P}}}
\newcommand{\pQa}{{\ensuremath{q_1}}}
\newcommand{\pQb}{{\ensuremath{q_2}}}
\newcommand{\dpr}[2]{{\ensuremath{(#1 \cdot #2)}}}

\preprint{CLNS 07/1997}       
\preprint{CLEO 07-06}         

\title{
Study of Di-Pion Transitions \\
Among $\Upsilon(3S)$, $\Upsilon(2S)$, and $\Upsilon(1S)$ States
}
\author{D.~Cronin-Hennessy}
\author{K.~Y.~Gao}
\author{J.~Hietala}
\author{Y.~Kubota}
\author{T.~Klein}
\author{B.~W.~Lang}
\author{R.~Poling}
\author{A.~W.~Scott}
\author{A.~Smith}
\author{P.~Zweber}
\affiliation{University of Minnesota, Minneapolis, Minnesota 55455}
% ...
\collaboration{CLEO Collaboration} 
\noaffiliation
%

\date{August 14 2007}

\begin{abstract} 
   We present measurements of 
   decay matrix elements for
   hadronic transitions of the form $\upsDec{n}{m}$, where $(n, m) =
   (3, 1), (2, 1), (3, 2)$.  We reconstruct charged and neutral pion
   modes with the final state Upsilon decaying to either $\mu^+\mu^-$ or
   $e^+e^-$.  Dalitz plot distributions for the twelve decay modes are
   fit individually as well as jointly assuming isospin symmetry,
   thereby measuring the matrix elements of the decay amplitude.  We
   observe and account for the anomaly previously noted 
   in the di-pion invariant mass distribution for the
   $\upsDec{3}{1}$ transition and obtain good descriptions of the
   dynamics of the decay using the most general decay amplitude
   allowed 
by partial conservation of the axial-vector current (PCAC)
considerations.  The fits
   further indicate that the \upsDec{2}{1} and \upsDec{3}{2}
   transitions also show the presence of 
   terms in the decay amplitude that were previously ignored,
although at a relatively suppressed level. 
\end{abstract}
\maketitle

%
\section{Introduction}

The transitions \upsDec{n}{m} are of particular interest as probes of
heavy quark and low energy QCD systems.  The large $b$~quark mass
causes the \bbBar bound state to have a very small radius 
($\sim 1 $~GeV$^{-1}$) and to be
non-relativistic ($(v/c)^{2}\approx 0.1$).  
This makes these transitions ideal to
study the process by which a pion pair is excited from the vacuum by
the gluon field.  The transitions 
among
the massive bound states
making up the \ups{n} family can be 
calculated
in terms
of multipole moments of the chromo-dynamic field, providing simple
relative rate and transition rule predictions.  The pion pair
excitation can be factored out and approximated separately.  Most
recent theoretical work has concentrated on this latter aspect of the
decays.

The \etaDec transition was the first decay of this form studied~\cite{London66},
followed some years later by the \psiDec 
transition~\cite{psiprime}.  The $\eta'$
decay is only barely above threshold, and so the transition cannot
show significant structure. 
Detailed study of the
kinematics confirmed this.  In contrast to this, the $\psi'$ decay has
decay dynamics very different from a phase space distribution.  The
di-pion invariant mass distribution of this decay shows strong
enhancement at larger values of $\mPiPi$.  However, this is consistent
with the presence of only the simplest term in the general Lorentz
invariant amplitude derived from 
PCAC
considerations \cite{Brown:1975dz,MBVoloshin:75}.  
This is supported by the isotropic
decay angular distribution of the pions, implying a minimal $D$-wave
component.

Previous CLEO data have been used to study 
\upsDec{n}{m}
transitions \cite{Butler:1993rq,Glenn:1998bd,Alexander:1998dq,Brock:pj},
with the \upsDec{2}{1} and \upsDec{3}{2} transitions following this same
pattern
in the di-pion invariant mass spectra as for the lighter mesons.
But the \upsDec{3}{1} transition has a second, strong rate enhancement near
the \diPi invariant mass threshold.  This enhancement and the
accompanying depletion at intermediate invariant mass 
are
inconsistent
with either pure phase space or the simple matrix element describing
the \psiDec observations.  
Either another term must be included in the
Lorentz invariant matrix element, or one must question the 
applicability of PCAC to the pion excitation and the validity of
the multipole expansion of the \bbBar bound state.

Various mechanisms have been suggested to explain this anomaly, such as
(i) large contributions from final state interactions~\cite{BDM,CKK},
(ii) a $\sigma$ isoscalar resonance in the $\pi\pi$ system~\cite{KII,Uehara},
(iii) exotic $\Upsilon -\pi$ resonances~\cite{VoloshinJTEP, BDM, ABSZ,GSCP},
(iv) an {\it ad hoc} constant term in the amplitude~\cite{Moxhay}, 
(v) coupled channel effects~\cite{LipkinThuan, ZK},
(vi) $S-D$ mixing~\cite{CKK2},
and (vii) relativistic corrections~\cite{Voloshin:2006}.

More recent 
experimental
analyses
with the very large
data sets accumulated by the $B$~factories at the $\ups{4}$ show
interesting behavior as well.  Belle~\cite{Abe:2005bi} and
$B$a$B$ar~\cite{BaBar4S2S1S} do not see such anomalous
behavior in the \upsDec{4}{1} transition, but 
$B$a$B$ar
does see such a double peaked 
structure in the
\upsDec{4}{2} transition.  

The shapes of the decay distributions originate in the details of the
excitation of the pion pair from the vacuum and the particular
projection of the initial state onto the final state.  Hence, the
enhancement of the decay rate at low $\mPiPi$, thus far considered an
anomaly, is a good probe of the details of low energy QCD in the
transitions of the bound states and the excitation of light hadrons
from the vacuum.

The general matrix element constrained by 
PCAC
was
derived by Brown and Cahn \cite{Brown:1975dz} and is further
constrained by treating the Upsilon transition 
as
a multipole
expansion as derived by Gottfried \cite{Gottfried:1977gp},
Yan \cite{Yan:1980uh}, Voloshin and Zakharov \cite{Voloshin:1980zf},
and others.  The general
transition amplitude is then given in non-relativistic form:
\begin{equation}
   \clg{M} = \clg{A} \dpr{\epp}{\epu} ( q^2 - 2 M_{\pi}^2 )\,+
             \clg{B} \dpr{\epp}{\epu} E_1 E_2 +
             \clg{C} ( (\epsilon'\cdot q_1) (\epsilon\cdot q_2) +
                       (\epsilon'\cdot q_2) (\epsilon\cdot q_1) )~,
   \label{Eq:BandCAmplitude}
\end{equation}
where $\epp$ and $\epu$ are the polarization vectors of the parent and
final state Upsilons, and $q_{1,2}$ are
the 
four-momenta
of the pions.  In the first term, $q^2$ is the
invariant mass of the pion pair.  The quantities $E_1$
and $E_2$ are the energies of the two pions in the parent rest frame,
essentially indistinguishable from the lab frame due to the large
masses of the Upsilons.\footnote {For transitions from the 
$\Upsilon$(3S), the parent frame and lab frame are 
virtually
identical.  Even for $\upsDec{2}{1}$ transitions, 
in which the $\Upsilon$(2S) comes from hadronic or electromagnetic
transitions from the $\Upsilon$(3S), the
parent's motion in the lab frame is unobservable other than in a small
broadening of recoil mass peak and a slight smearing of reconstructed
variables.}  
The third, or ``${\cal C}$'' term in this
expression couples transitions via the chromo-magnetic moment of the
bound state $b$~quarks, hence requiring a spin flip. This is expected
to be 
highly
suppressed by the large mass of the $b$~quark, so we expect only the
first two terms to contribute.
Neglecting the dependence on the parent and final state
Upsilon polarizations (which apply only to the 
${\cal C}$-term), we have only
two degrees of freedom, the Dalitz variables $q^2 = M^2_{\pi\pi}$ and
$r^2 = M^2_{\Upsilon\pi}$.  
In writing this amplitude, we have assumed the chiral limit, so that 
a fourth term, $gM_{\pi}^{2}$, is taken to be 
zero~\cite{MannelUrech, newVoloshin}.

The
expression in Eqn.~\ref{Eq:BandCAmplitude} can be made
fully Lorentz invariant by rewriting the energy product in the ${\cal B}$ 
term as
\begin{equation}
   E_1 E_2 \approx [(P^{\prime}\cdot q_{1})(P\cdot q_{2}) + (P^{\prime}\cdot q_{2})(P\cdot q_{1})]/[2M_{\Upsilon '}M_{\Upsilon}],
   \label{Eq:E1E2expansion}
\end{equation}
with $\pPp$ and $\pPu$ being  the initial state and final state $\Upsilon$
four-momenta.

The quantities $\clg{A}$, $\clg{B}$, and $\clg{C}$ are form factors
that depend on the detailed dynamics of the decay.  They are in
principle functions of the Dalitz variables $q^2$ and $r^2$.  However,
we expect them to vary on the scale of $\LambdaQCD$, which is
comparable to the total energy release of the decays, so to first
order we assume they are complex constants.  Angular structure or $\mPiPi$
dependence beyond that indicated in the explicit amplitude,
Eqn.~\ref{Eq:BandCAmplitude}, would be an indication of the
non-constancy of these form factors, or alternately the breakdown of
the assumptions leading to Eqn.~\ref{Eq:BandCAmplitude}.

\begin{figure*}
   \centerline{
      \resizebox{0.80\textwidth}{!}{
       \includegraphics{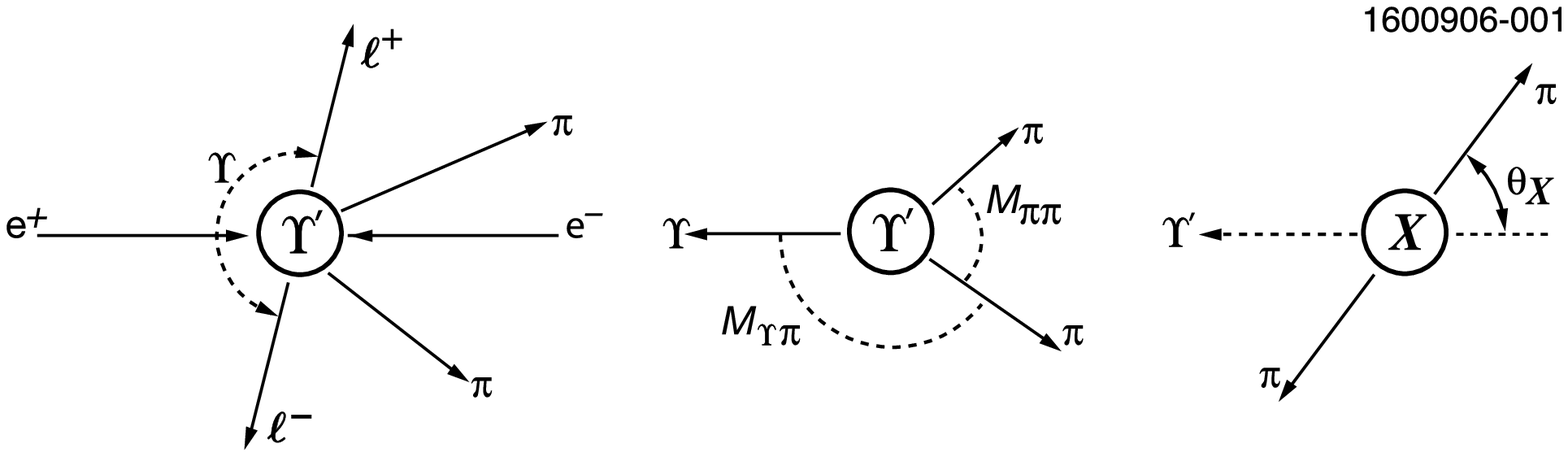}}
   }
   \caption[The decay process \upsDec{n}{m}]
           { (Left) The decay \upsDec{n}{m} follows the production
             of an initial state labeled $\Upsilon'$ which decays to
             an $\Upsilon \pi \pi$ state.  In our analysis, the final state $\Upsilon$
             decays to a lepton pair 
	     whose momentum vectors are 
	     very nearly back-to-back
             due to the large energy release.  (Center) The decay of
             the initial state Upsilon is governed by two kinematic
             variables, the Dalitz masses $M_{\Upsilon\pi}$ and
             $M_{\pi\pi}$.  (Right) Alternately one can think of the
             $\pi \pi$ system as a composite, $X$, and study its structure
             via the pion ``decay'' angles.}
   \label{fig:upsDec}
\end{figure*}

The di-pion transition can be interpreted as taking place in
sequential two-body decays through a fictitious intermediate state $X$
via the chain $\upsXDec{n}{m}$ and $X \rightarrow \pi \pi$ (see 
Fig.~\ref{fig:upsDec}).  In this
view we can define the helicity angle of the $X \rightarrow \pi \pi$
decay in the usual manner of the Jacob and Wick formalism.  The polar
helicity angle is referred to as $\theta_{X}$.  
Its cosine is used interchangeably
with the second Dalitz variable, $r^2$, as they are linearly related:
\begin{equation}
   2 r^2 = 2 M_\pi^2 + M_{\Upsilon'}^2 + M_{\Upsilon}^2 - q^2
           - \cthx \sqrt{ \frac{1}{q^2}
                          \left( q^2 - 4 M_\pi^2 \right)
                          \Lambda_3( M_{\Upsilon'}^2,
                                     M_{\Upsilon}^2, q^2 ) }
\end{equation}
where $\Lambda_3( a, b, c ) = a^2 + b^2 + c^2 - 2 a b - 2 a c - 2 b
c$.  
These variables ($r^{2}$ or $\cthx$) carry structure 
from the second term in the
amplitude due to the following relation:
\begin{equation}
   E_1 E_2 = \frac{1}{4}\left( ( E_1 + E_2 )^2 - \Delta E^{2}_{\rm max}
             \cos^2 \theta_X \right)~,
\end{equation}
with $\Delta E \equiv E_{2} - E_{1}$.
Because the initial state and final state Upsilons are essentially at rest, the
energy sum $E_1 + E_2$ is 
nearly
a constant and equal to the mass
difference between the Upsilons.  
For the $\pi^{+}\pi^{-}$ final state, $\theta_{X}$ is defined as the
angle of the 
positive pion, with
$-1 < \cos \theta_X < 1$; for the $\pi^{0}\pi^{0}$
final state, because
one cannot distinguish between the two neutral pions, we take
$0 < \cos \theta_X < 1$.

Finding the presence of a non-zero $\clg{C}$ term would indicate the
breakdown of the multipole expansion, {\it i.e.,} of the
assumption that the pion pair excitation is independent of the Upsilon
transition process from $n^3S_1$ state to 
$m^3S_1$, and that the spin flip of the $b$ quarks is
suppressed.  However, finding a non-zero $\clg{C}$ term 
could also be due to distortions of the
distribution not accountable for by using only the first two 
terms with complex, but {\it constant}, coefficients ${\cal A}$ and ${\cal B}$.

\section{Data Sets and Event Selection}

Data were collected with the CLEO~III detector which is described in
detail
elsewhere~\cite{Viehhauser:2001ue,Peterson:2002sk,Artuso:2005dc}.  
In this analysis we observe
$e^\pm$, $\mu^\pm$, $\pi^\pm$, and $\gamma$
particles in the final state, and so 
use
both the tracking and
calorimetry information from the detector, as well as lepton
identification.  Thus we 
employ 
global event, track, lepton,
shower, and neutral pion selection criteria, in addition to signal and
background identification criteria.

The data were taken while running
on the $\ups{3}$ resonance, subject to standard CLEO data quality
selections,
and represent an integrated luminosity of
\nUnits{1.14}{\invfb}, 
and an $\ups{3}$ production yield of
$(\nError{4.98}{0.01})\times10^6$.  The $\ups{2}$ sample is
obtained by reconstruction of sequential decays, $\ups{3} \rightarrow
\ups{2}\,+\,{\rm anything}$, occurring in this sample.  The $\ups{2}$
population of $(\nError{5.27}{0.40})\times10^5$
is estimated from the branching fraction \cite{PDG:2006} of
$\nError{10.6\%}{0.8\%}$ for the decay $\ups{3} \rightarrow
\ups{2}\,+\,{\rm anything}$, which is dominated by pion pair transitions and
sequential photon decays through the $\chi_b$(2P) states. 

All integrals needed in the analysis (for evaluation of acceptances
and efficiencies) are calculated via the Monte Carlo method.  Physics
event generation is performed 
using the Lund Monte
Carlo~\cite{Sjostrand:2001yu} embedded in the CLEO physics Monte Carlo
{\tt QQ}~\cite{QQ}.  The Lund event generator is used because it accurately accounts for the
physics of the QCD bound state production.  The $\ups{3}$ produced in
the $e^+e^-$ collision is then decayed according to 
standard
decay
tables and the detector response to the decay products is simulated
using 
the physics simulation package
{\tt GEANT}~\cite{CERNLib:GEANT}.

In general, since all integrals are performed with respect to the
natural measure over phase space, only phase space decays need be
simulated.  The decay amplitude is known exactly as a function of the
decay kinematics, so all inputs to the matrix element extraction
(other than acceptance and efficiency) are known to the precision of
detector reconstruction.

\begin{figure*}
   \centerline{
      \resizebox{0.66\textwidth}{!}{
         \includegraphics{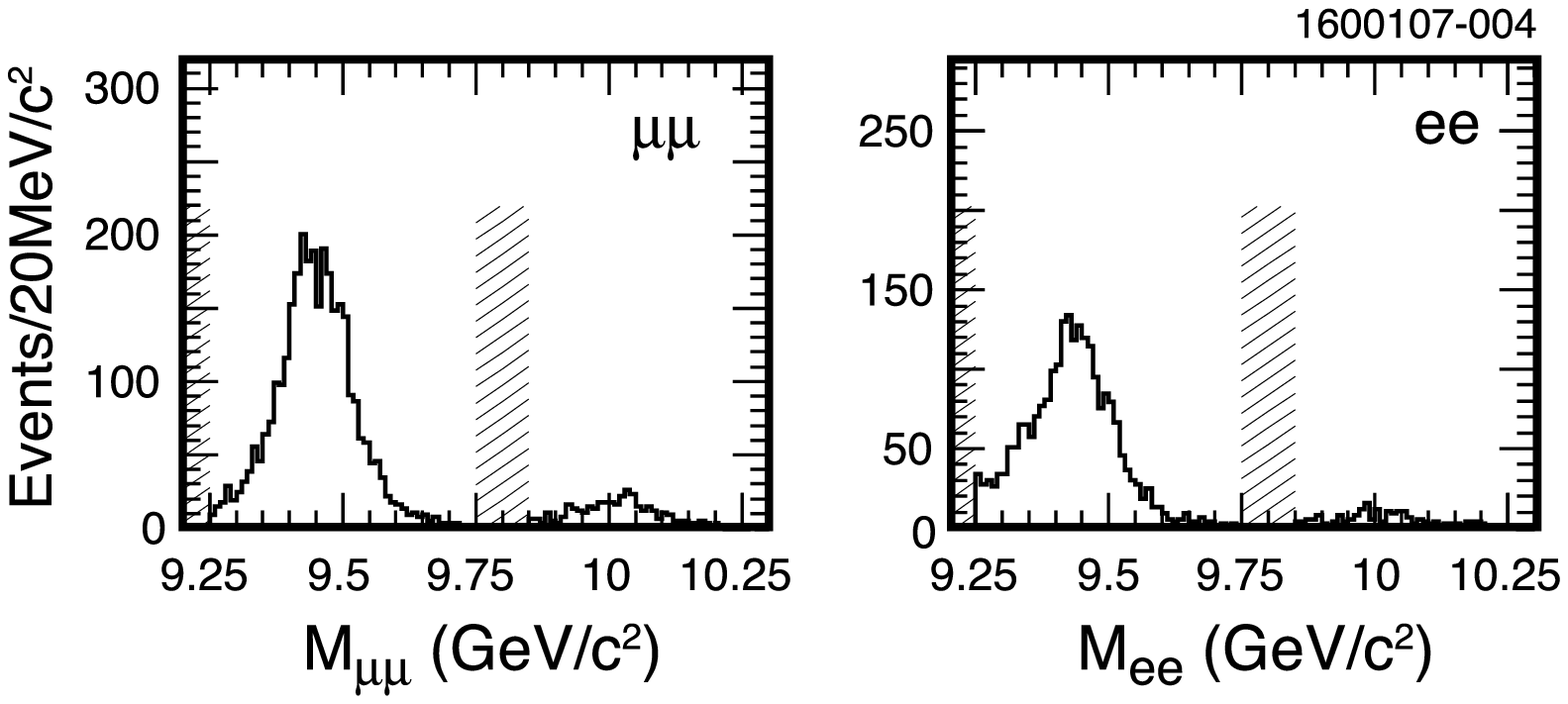}}
   }
   \caption[Di-lepton mass spectra, raw and with signal selections] {
           Di-lepton invariant mass distributions for lepton
           pairs; the abscissa is the di-lepton invariant mass,
           showing peaks at the masses of the \ups{1} and \ups{2}
           mesons.  The hatching indicates the limits to the
           invariant mass selection windows.
           Candidates are
           plotted after the signal selection described in 
	   Section II.D.  At
           left are the di-muon candidates and at right the
           di-electron candidates.}
   \label{fig:lepMassSpectra}
\end{figure*}

\begin{figure*}
   \centerline{
      \resizebox{0.66\textwidth}{!}{
         \includegraphics{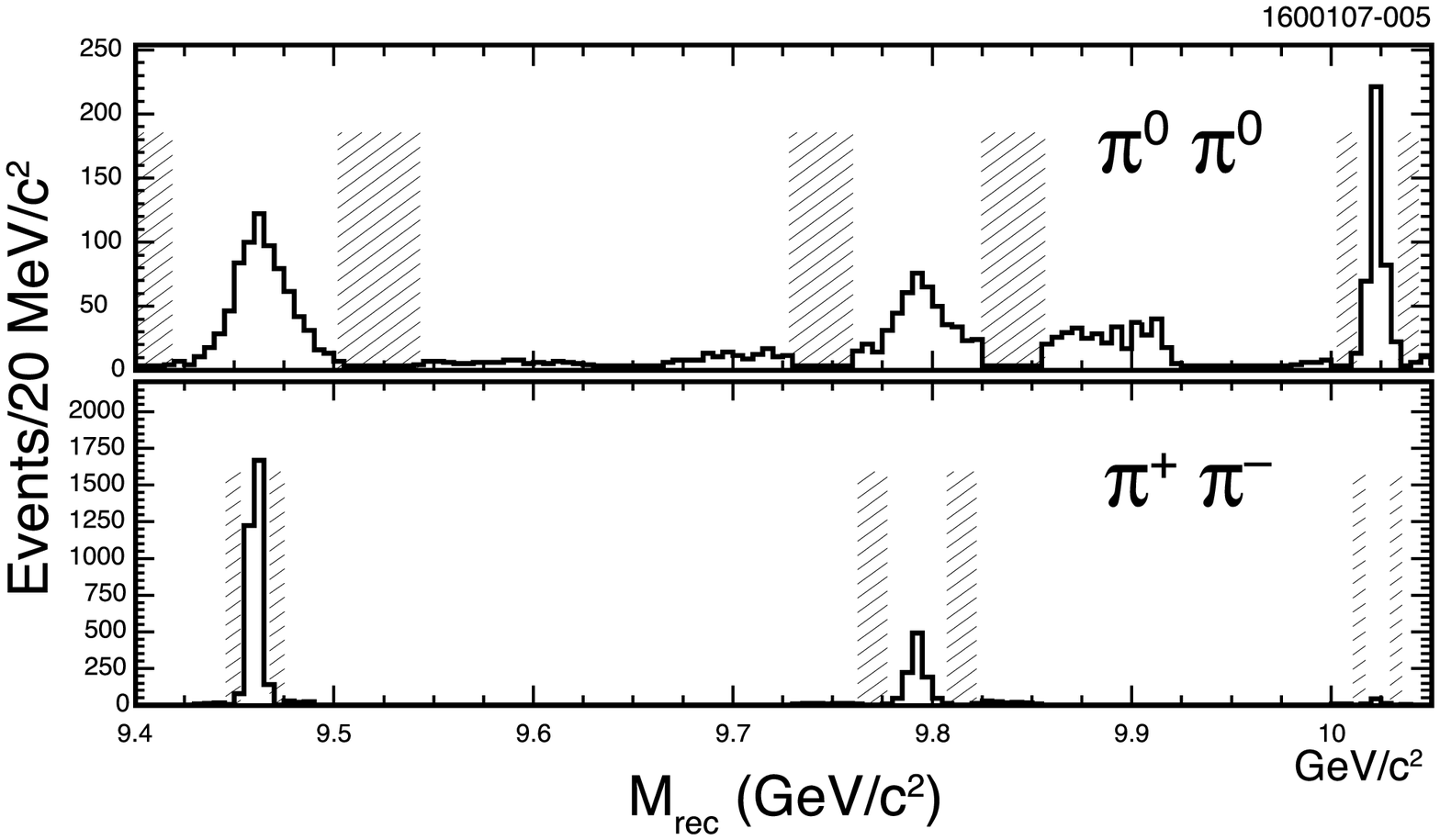}}
   }
   \caption[Recoil mass spectra, raw and with signal selections] {
           Recoil mass, $M_{\rm rec}$, distributions for all modes.
           The upper plot is generated from neutral decays,
           $\upsNDec{n}{m}$ and the lower from charged decays,
           $\upsCDec{n}{m}$.  The final signal selections
           (track quality, pion quality, di-lepton mass, {\it etc.}) have
	   been applied.  The peaks at the $\Upsilon$(1S) and
	   $\Upsilon$(2S) masses correspond to decays to these resonances
	   from an $\Upsilon$(3S) parent.  The peaks at 9.8 GeV/$c^{2}$ are
           from $\Upsilon$(2S)$ \to\Upsilon$(1S)$\pi\pi$ decays.
           The hatching shows the bounds on the recoil
	   mass values for the three transitions.
	   See also the window
	   definitions in Tab.~\ref{tab:RecMassPeaks}.  Yields are set to zero in
           the regions that correspond neither to signal nor to 
	   sidebands.}
   \label{fig:recMassSpectra}
\end{figure*}

We select events containing two leptons ($\mu^+\mu^-$ or $e^+e^-$) and
two pions ($\pi^+\pi^-$ or $\pi^0\pi^0$).  
All low momentum tracks are assumed to be pions, because
there is insufficient phase space for the production of
a pair of kaons in a transition among any two of the
three bound state Upsilons.
Electrons and muons are identified by their energy loss and
penetration depth in the detector as detailed below, and are required
to be consistent with originating from either an \ups{2} or an \ups{1}
decay.  The pion candidates are constrained to come from a common point
at the beam location and the recoil mass 
($M_{\rm rec}^2 = P_{\rm rec}\cdot P_{\rm rec}; P_{\rm rec} \equiv P_{\rm
beam} - q_1 - q_2$; see below) is used to identify the transition.  The
lepton pair invariant mass spectra and the recoil mass spectra are
shown in Figs.~\ref{fig:lepMassSpectra}
and~\ref{fig:recMassSpectra}, respectively.

\subsection{Global Event Selection}

The data used in this analysis are required to have been
taken while running on the \ups{3} resonance energy.  
Global event characteristics are used to preselect the events.
Excessive tracks or showers in an event can dramatically increase the
combinatoric background.  To avoid this, reconstructed events are
selected subject to upper limits on number of charged particle tracks
and number of calorimeter showers.  To establish 
conservative
limits, signal
Monte Carlo is studied for  \upsDec{2}{1}
transitions, which are the ``worst case'',  in that extra 
tracks and showers in these modes arise from the
initial transition from the \ups{3} to the \ups{2}.  Neglecting stray
particles and secondary showers, there should be no more than four low
momentum charged particle tracks and no more than eight
electromagnetic showers in signal events.  Comparison between data and
Monte Carlo show good agreement in the number of tracks and showers
found in the selected events.

\subsection{Selection of Final State Particles}

All
candidate charged tracks are required to satisfy 
quality criteria.  
They must:
\begin{list}{}{\leftmargin36.0pt \partopsep0.0pt
    \parskip0.0pt \topsep2.0pt \itemindent0.0pt \parsep2.0pt
    \itemsep0.0pt \labelsep2.0pt }
\item[-]{come from within \nUnits{5}{\cm} of the origin along the beam
  axis (detector $\hat{z}$ axis)};
\item[-]{come within \nUnits{5}{\mm} of the beam axis (impact
  parameter)};
\item[-]{have momentum less than the beam energy;}
\item[-]{have a good helix track fit, with $\chi^2$ per hit less than 20}.
\end{list}
These requirements are applied to all track candidates and are
augmented with identification criteria for leptons (see below) before
being accepted as decay candidates.

The charged transition pions frequently are of such low transverse
momentum that they make two or more semi-circular arcs in
the tracking volume.  These ``excess'' tracks  are removed by 
comparing the helix 
parameters, taking into account
the expected energy loss as these pions spiral through the
drift chamber.

Candidate muons and
electrons are required to have high momentum by requiring their 
transverse momentum to be 
$p_{T} > 1$ GeV/$c$,
which removes a large fraction of the events with non-leptonic
Upsilon decays.  
Because
the leptons we seek originate from the decay of
objects 
more massive
than \nUnits{9.4}{\mGeV}, they pass this requirement
easily.

Muons are selected from among good tracks and 
are
additionally required to
penetrate the muon chambers to a depth of at least 
three
interaction
lengths.  The ratio of energy deposition in the calorimeter to track
momentum must also be less than one half, $E/pc < 0.5$.

Electrons are selected from among good tracks and are additionally
required to have a ratio of energy deposited in the calorimeter to
track momentum $E/pc > 0.5$, as well as having 
a profile of energy deposition consistent with that of an 
electromagnetic shower
and a good spatial match between the shower and the
track.  The $E/pc$ ratio selection is a very loose requirement added
only as a 
precaution
against muons contaminating the electron sample.

The di-lepton mass is loosely required to
be that of the final state Upsilon being studied, as shown in
Fig.~\ref{fig:lepMassSpectra}.  For the $\Upsilon$(1S) we require
$9.25 < M_{\ell\ell} < 9.75$ GeV/$c^{2}$, while for the
$\Upsilon$(2S) we demand $M_{\ell\ell} > 9.85$ GeV/$c^{2}$.
Due to the large widths of these invariant mass peaks, no 
side band selection is
performed in this variable, but rather only in the recoil
mass distribution.  

The $\pi^0$ candidates are reconstructed from photon
pairs.  This begins by applying selection criteria to the showers.
To be considered a photon, a shower must:
\begin{list}{}{\leftmargin36.0pt \partopsep0.0pt
    \parskip0.0pt \topsep2.0pt \itemindent0.0pt \parsep2.0pt
    \itemsep0.0pt \labelsep2.0pt }
\item[-]{have energy greater than \nUnits{30}{\eMeV}};
\item[-]{have a lateral shower profile consistent with that of a photon};
\item[-]{be inconsistent with the extrapolation of any track in the
  detector};
\item[-]{not include noisy channels in the calorimeter};
\item[-]{not be in the overlap region between the barrel and endcap
calorimeter modules};
\item[-]{not be in the ring of crystals closest to the beam axis}.
\end{list}

Showers satisfying these selection criteria are considered to be
photons and are combined into $\pi^0$ candidates.  Photon pairs are
required to have an invariant mass 
within \nUnits{50}{\mMeV} of the
nominal $\pi^0$ mass, $M_{\pi^{0}}$.  
They are then required to fall within the
asymmetric window
\begin{equation}
   -4 < \frac{M_{\gamma \gamma} - M_{\pi^0}}{\sigma_{\gamma \gamma}} < 3~.
\end{equation}
The photon-pair mass resolution, $\sigma_{\gamma \gamma}$, is typically
5-7 MeV/$c^{2}$.
Candidate photon pairs are then 
kinematically constrained (subject to the measured
uncertainties on energies and shower spatial locations) to have an
invariant mass equal to 
$M_{\pi^{0}}$.
To be used,
$\pi^0$ candidates are further required to have a
successful kinematic fit with confidence level (one degree of freedom) 
greater than $0.1\%$.

\subsection{Recoil Mass and Signal and Background Regions}

We select events for each transition by cutting on the mass of the
system 
recoiling against the two pions in
the 
$\Upsilon^{\prime} \rightarrow\pi\pi + $''anything'' 
decay:
$M^2_{\rm rec} = M^2_{\Upsilon'} + q^2 -2 q\cdot P'$, where, as above, 
$q = q_{1} + q_{2}$ and $P'$ is the Lorentz momentum of
the initial state Upsilon.  
Given the large mass of the initial state Upsilon, the
dot product simplifies and the recoil mass can be well approximated by
$M^2_{\rm rec} \approx M^2_{\Upsilon'} + q^2 - 2 M_{\Upsilon'} (E_1 + E_2)$.
For the cascade decays, \upsDec{2}{1}, this is not quite correct
because the Lorentz momentum of the initial state Upsilon (the \ups{2}) is
not equal to the beam momentum.  However, because the total momentum of
the pions is small and the initial state is approximately at rest, using the
incorrect momentum for the initial state does not significantly change the
recoil mass distribution other than to shift it by the difference
between \ups{3} and \ups{2} masses.  Hence, we expect to find three
recoil mass peaks.  The transitions originating from the \ups{3} will
generate recoil mass ($M_{\rm rec}$) peaks at the masses of the
\ups{1} and \ups{2}, while the \upsDec{2}{1} decays will yield a peak
at \nUnits{9.79}{\mGeV}.  These three peaks are clearly visible in
Fig.~\ref{fig:recMassSpectra}.

The recoil mass, $M_{\rm rec}$, is measured rather accurately, especially in the charged case, due to
the good resolution on the momenta of the low-momentum pions.  
It is still quite good for the
neutral modes where the total pion momentum is given as the sum of
momenta of two $\pi^0$ candidates reconstructed from the calorimeter
showers.  

\subsection{Signal and Background Selection}

The fit requires signal and background samples.  They are determined
as a function of the recoil mass, $M_{\rm rec}$, only.  The recoil
mass peak widths are determined from Monte Carlo with tight selections
on variables other than the recoil mass.  
These widths
are then used to
determine mass windows to select events in both Monte Carlo and data
samples.  The signal regions are defined as the range within three
times the peak width of the nominal recoil mass, while the backgrounds
are the regions from six to twelve times the peak width from the nominal
mass above and below the peak mass.  The masses and widths used to
define these regions are listed in Table~\ref{tab:RecMassPeaks}.  The
width of the recoil mass distribution in the decays \upsCDec{2}{1} is
roughly twice that of the direct decays.  This is due to the boost of
the initial state Upsilon imparted in its production by the cascade from the
\ups{3}.  The edges of the signal windows are indicated by the
hatching in Fig.~\ref{fig:recMassSpectra}.  Note that 
in Fig.~\ref{fig:recMassSpectra} the yield in the regions 
not used for {\it either} 
signal or background definition have been set to zero.

\begin{figure*}
   \centerline{
	\resizebox{0.90\textwidth}{!}{\includegraphics{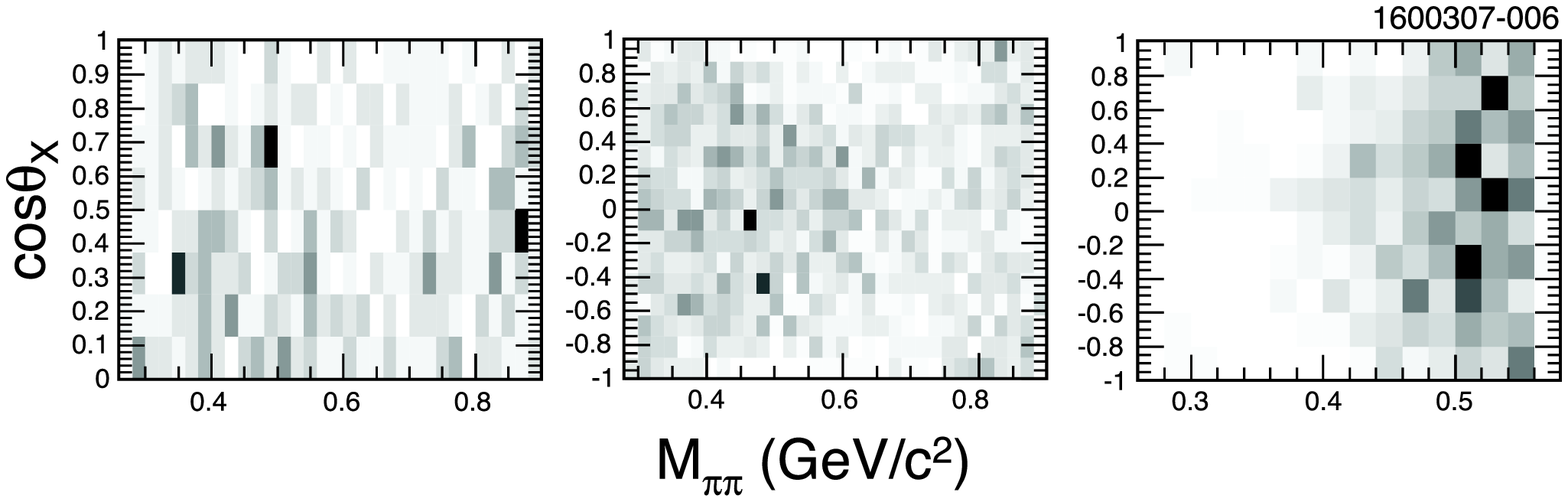}}
	}   
   \caption{ 
Candidate events that
passed all selection criteria,
and that have
the final state Upsilon decaying to $\mu^{+}\mu^{-}$.  In the middle
is the decay $\upsCDec{3}{1}$.  To the left is its neutral counterpart
$\upsNDec{3}{1}$.  
To the right is the charged transition
$\upsCDec{2}{1}$, with the bulk of its distribution at large values of
dipion invariant mass.  In each plot there are ten degrees of grey-scale 
ranging from white (lowest occupancy per bin) to black (highest occupancy).}
   \label{fig:datax3mm}
\end{figure*}

\begin{figure*}
   \centerline{
     \resizebox{0.90\textwidth}{!}{\includegraphics{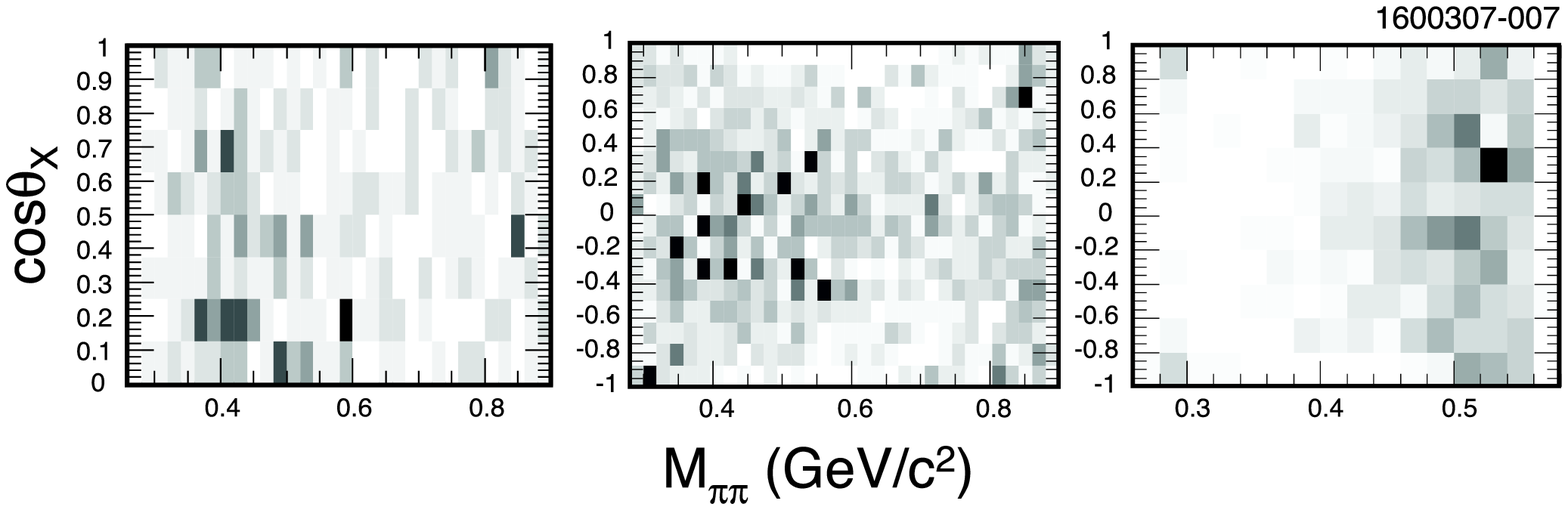}}
    }   
   \caption{ 
Candidate events that have
passed all selection criteria,
and that have
the final state Upsilon decaying to $e^{+}e^{-}$.  As in the prior
plot, three transitions are, left to right, $\upsNDec{3}{1}$, $\upsCDec{3}{1}$,
and $\upsCDec{2}{1}$.}
   \label{fig:datax3ee}
\end{figure*}

The Dalitz plot distributions for the selected data in
six
of the twelve final states are shown in 
Figs.~\ref{fig:datax3mm} and ~\ref{fig:datax3ee}.
Comparison of
the $\pi^{0}\pi^{0}$
and  $\pi^{+}\pi^{-}$ for the  \upsDec{3}{1} shows the depletion
in charged particle efficiency at moderate di-pion invariant mass and
large $|{\rm cos}~\theta_{X}|$.  Comparison of the charged modes for
\upsCDec{3}{1} and \upsCDec{2}{1} shows, in two dimensions, the obvious
disparity between the two distributions.


\begin{table*}
   \centerline{
      \begin{tabular}{lcccc} \hline \hline
         Transition  &  Recoil Mass &
                        Width (Data) &  Width (MC)  &
                        Width (Cut) \\
                     &  (\mMeV)  &  (\mMeV)  &  (\mMeV)  &  (\mMeV)  \\
         \hline
         \upsCDec{3}{1}  &  9~460.4 &  2.4 &  2.5 &  2.5 \\
         \upsCDec{2}{1}  &  9~792.4 &  5.0 &  5.0 &  5.0 \\
         \upsCDec{3}{2}  & 10~023.3 &  2.2 &  1.9 &  2.1 \\
         \hline
         \upsNDec{3}{1}  &  9~460.4 & 15.0 & 12.7 & 13.8 \\
         \upsNDec{2}{1}  &  9~792.4 & 10.9 & 10.5 & 10.7 \\
         \upsNDec{3}{2}  & 10~023.3 &  3.4 &  3.4 &  3.4 \\
         \hline \hline
      \end{tabular}
   }
   \caption{ Recoil mass distribution central values and widths for
             the signal and background selections used in the fit.
             The central values and the widths agree well between data
             and Monte Carlo.  The signal windows are defined as the
             region within three times the cut width (last column) of
             the central mass and the background windows are defined
             as the region from six to twelve cut widths from the
             center on either side.  The background subtraction is
             only important for the cascade decays for which there is
             a large contribution to the signal region from event
             combinatorics.  }
   \label{tab:RecMassPeaks}
\end{table*}

\section{Matrix Element Fits}

\subsection{The Likelihood Fitter}

The binned likelihood fit to the kinematic distributions of the
$\Upsilon$(mS)$\rightarrow\Upsilon$(ns)$\pi\pi$ decays is designed to deal
correctly with the low bin yields expected from dividing approximately
2000 events over a two 
dimensional space with 
more than ten
bins per dimension.  The general case of this problem is solved in
Ref.~\cite{Barlow:dm}.  Specific details of our application of this technique,
including notes on variable smearing and background inclusion, are
found in the Appendix.
We fit the decay distributions to a product of the squared modulus of
the decay amplitude and the phase space density sculpted by the
detector acceptance.  The matrix element has a known analytical form
(see Eqn. \ref{Eq:BandCAmplitude})
as a function of the form factors $\clg{A}$, $\clg{B}$, and $\clg{C}$,
which are taken as complex constants.
Its leading angular structure is known, and so long as the form
factors are known, too, the entire amplitude can be described exactly.
However, we cannot model the detector acceptance in analytic form,
so we approximate its effect via Monte Carlo integration.  

\begin{figure*}
   \centerline{
     \hspace*{\fill}
     \resizebox{0.40\textwidth}{!}{\includegraphics{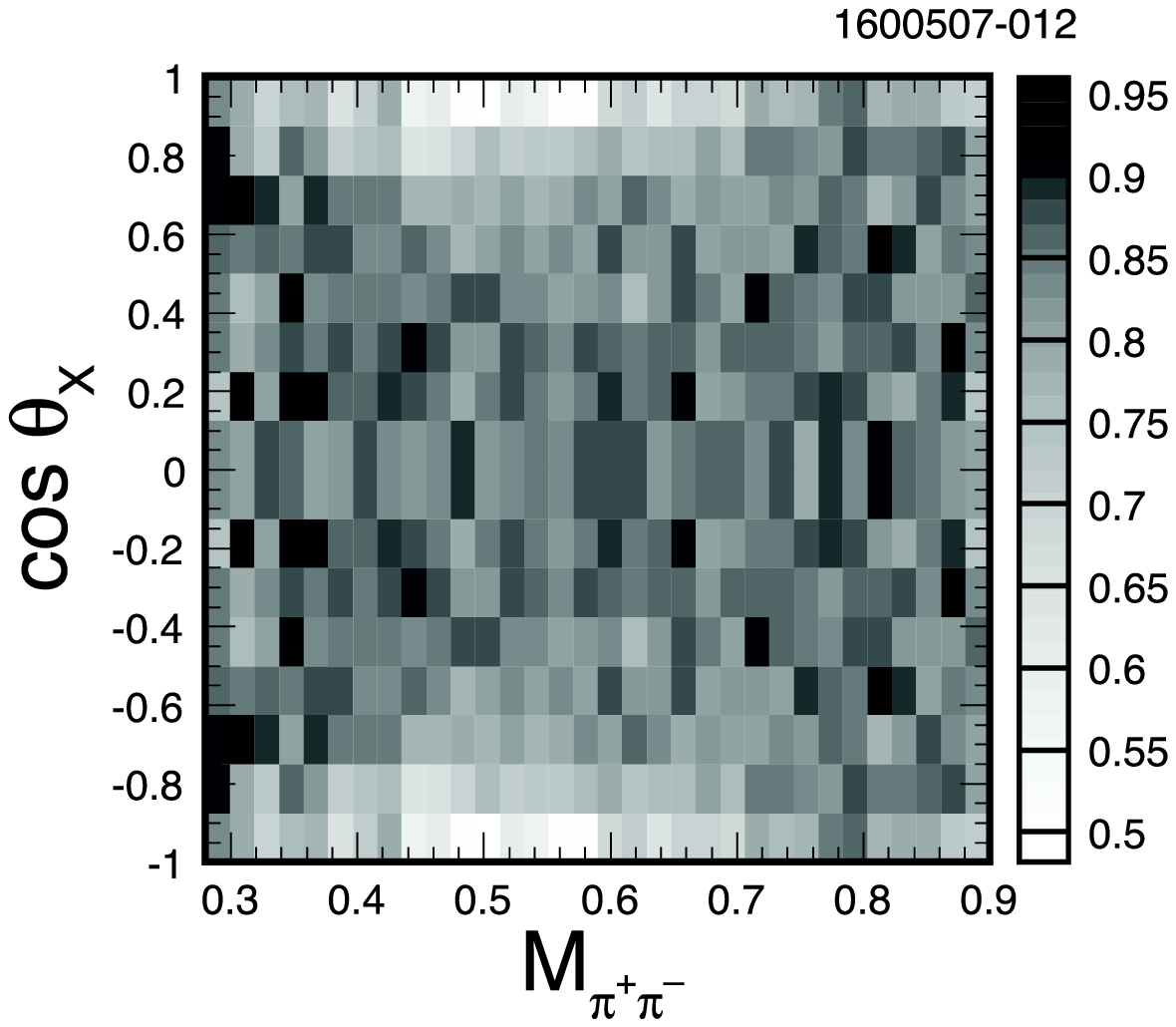}}
     \hspace*{\fill}
    \resizebox{0.40\textwidth}{!}{\includegraphics{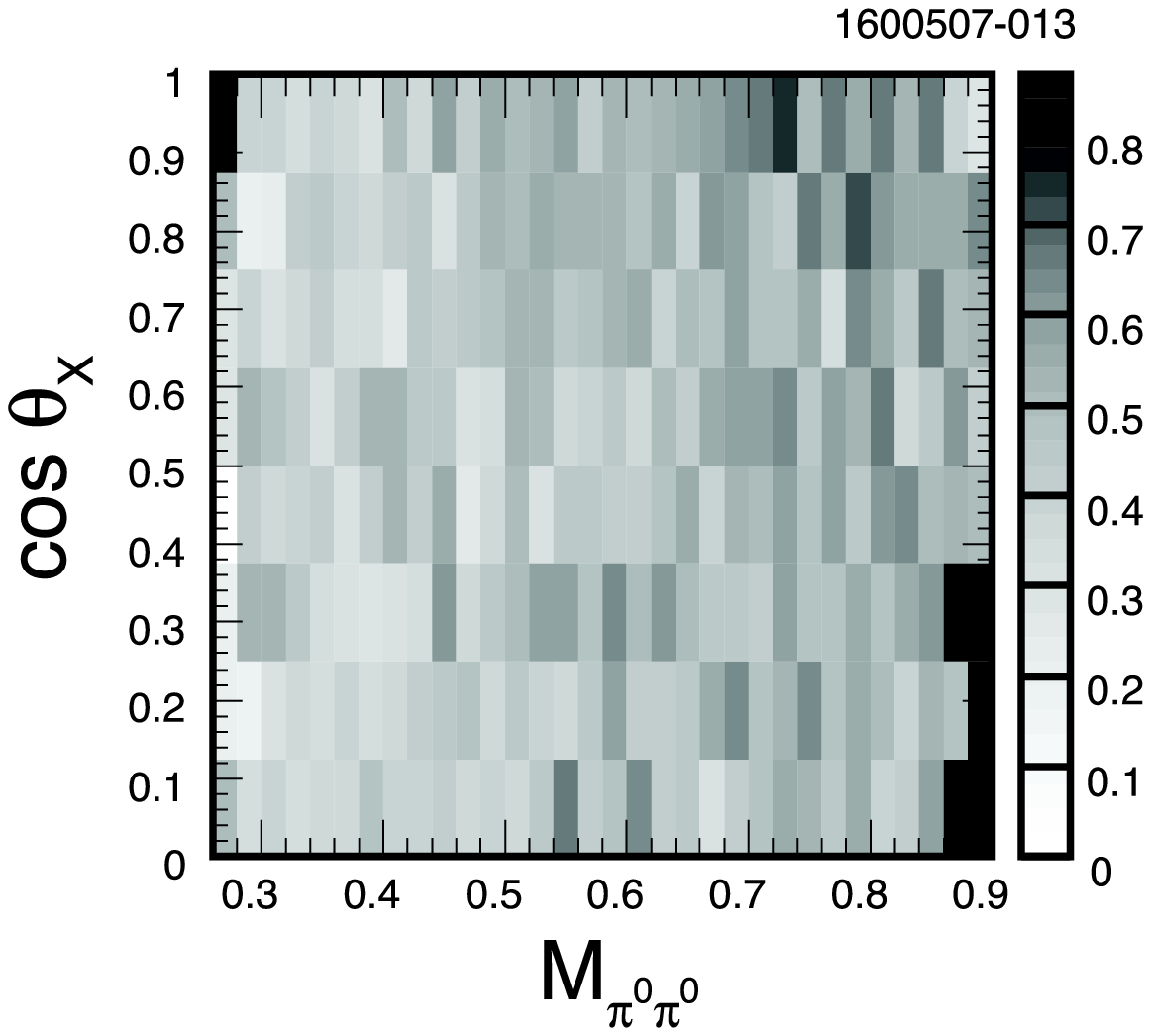}}
    }   
   \caption{ The efficiency-sculpted phase space 
in the two-dimensional plane for
the transitions  
$\Upsilon$(3S)$\to\Upsilon$(1S)$\pi^{+}\pi^{-}$ (left) and 
$\Upsilon$(3S)$\to\Upsilon$(1S)$\pi^{0}\pi^{0}$ (right).  Note that the
neutral final state has a more uniform efficiency, especially in the region
of moderate di-pion mass and large $|{\rm cos}(\theta_X)|$. }
   \label{fig:eff}
\end{figure*}

We determine the integral of the phase space density in a bin in 
($q^{2}, {\rm cos}~\theta_{X}$),
sculpted by acceptance and efficiency,
by counting Monte Carlo
events that pass the selection criteria and fall into that bin. 
In Fig.~\ref{fig:eff} we show the two-dimensional phase space
after such sculpting.  Note that while the overall efficiency for
the neutral final state is lower than for its charged counterpart,
the former is more uniform, particularly in the regions of intermediate
$M_{\pi\pi}$ and large $|{\rm cos}~\theta_{X}|$.  
For
each bin of the observed distribution we predict the number of events
as a function of the matrix element parameters by multiplying the
Monte Carlo integral for that bin by the exactly calculated matrix
element value for that bin.  This approach avoids generating Monte
Carlo integrated templates for each component of the angular
distribution and reduces the uncertainty due to finite Monte Carlo
sample size.

\begin{figure*}
   \centerline{
     \resizebox{0.90\textwidth}{!}{\includegraphics{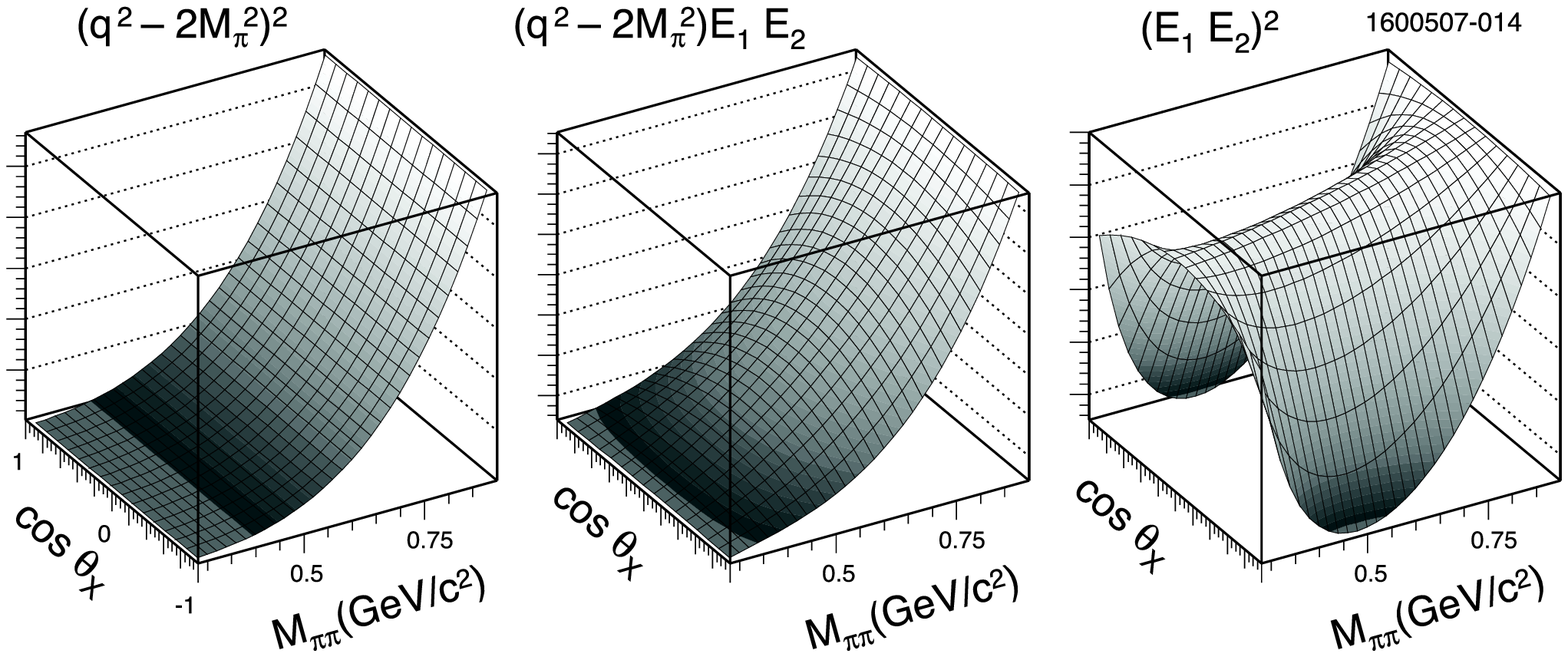}}
    }   
   \caption{ The three functions used in the fit for the $\Upsilon$(3S)
decay to $\Upsilon$(1S)$\pi\pi$.  From left to right these 
are for the pure ${\cal A}$ term, the interference term, and the 
pure ${\cal B}$ term. }
   \label{fig:ME}
\end{figure*}

To fit the decay distribution we take the squared modulus of the decay
amplitude, Eqn.~\ref{Eq:BandCAmplitude}, and decompose it as a sum
of six functional forms each multiplied by one of $|\clg{A}|^2$,
$|\clg{B}|^2$, $|\clg{C}|^2$, $\Re(\clg{A}^*\clg{B})$,
$\Re(\clg{A}^*\clg{C})$, or $\Re(\clg{B}^*\clg{C})$.  For
normalization, the
matrix element $\clg{A}$ is set to unity.

The functional forms (\eg $(q^2 - 2 M_\pi^2)^2$) depend on the Dalitz
variables and are pre-evaluated into templates over the Dalitz space.
The fitter then seeks the best fit as a function of the matrix 
element ratios
$\clg{A}$, $\clg{B}$, and
$\clg{C}$.  The input to the fitter consists of only the data,
background, and phase space Monte Carlo binned across the Dalitz plot,
and the component templates of the decay distribution derived from the
exact decay amplitude, but taking into account the kinematic smearing
and acceptance and efficiency effects due to reconstruction as
determined from the detector simulation.
The backgorund component is scaled by the ratio of the signal region width
(6 $\sigma$; see Section II.D) to the total backgorund sideband width
(nominally 12$\sigma$).

In Fig.~\ref{fig:ME}
we show the functional forms for $|\clg{A}|^2$, $\Re(\clg{A}^*\clg{B})$,
and $|\clg{B}|^2$ for the case of 
$\Upsilon$(3S)$\to\Upsilon$(1S)$\pi\pi$.
In our experiment, the 
complementarity of the neutral and charged final states is 
particularly important in that the 
rightmost of these (the form for $|\clg{B}|^2)$)
depletes the region for which the $\pi^{+}\pi^{-}$ channel
has falling efficiency.
Consistent results between the $\pi^{0}\pi^{0}$ and $\pi^{+}\pi^{-}$ 
transitions
gives us confidence that the simulation of this fall-off in efficiency 
is reliable.
The matrix element extraction procedure is tested ``end-to-end'' 
by simulating signal
with known matrix elements in Monte Carlo and comparing the fit result
and its uncertainty with the known inputs.  Samples of the same size
as the observed yield are generated and fit identically to the data.
The results yield standard normal distributions in the observed
uncertainty scaled residuals for widely distributed seed matrix
element values.  This confirms the fitter is unbiased at the level of
precision to be expected from the sample size of the measurement.

\subsection{Fits with ${\cal C} = 0$}

The fits to the two dimensional distributions of $\mPiPi$ and $\cthx$
determine the matrix element $\BoAf$ and $\CoAf$.  The extracted
values of $\Re(\BoAf)$ and $\Im(\BoAf)$ are summarized in
Table~\ref{tab:sep_sim_fit}, subject to the constraint that $\clg{C}
\equiv 0$.  In that we only measure the cosine of the phase
difference between ${\cal B}$ and ${\cal A}$, $\Im(\BoAf)$ is only 
known to within a sign.  
The upper set of matrix elements are obtained from
independent fits to 
ten individual decay modes; we cannot
individually fit the two modes associated with
$\Upsilon$(3S)$\to\Upsilon$(2S)$\pi^{+}\pi^{-}$ because of their
limited statistics. The lower
set of three are from the simultaneous fits of all final states for
each given Upsilon transition.

In the simultaneous fits the relative branching ratios between modes
are not constrained, but it is assumed that the di-pion excitation
dynamics is independent of the charge of the pion final state
(isospin symmetry) and thus the decay distributions should
be identical to within statistical fluctuations for all transitions
between the same Upsilon states.  This assumption is supported by the
consistency among the matrix element values extracted independently,
as well as their consistency with the value extracted from the
simultaneous fit.  In particular, the four final states studied for the
transition from $\Upsilon$(3S) to $\Upsilon$(1S) show excellent
agreement between the two lepton species and between charged and
neutral
pions.

\begin{table*}
   \centerline{
\begin{tabular}{llcc}
\hline \hline
Individual Fits   &                   & \ReBA                  & \ImBA~                \\
\hline
\upsCDec{3}{1};   & \uuuDec           & \nError{-2.514}{0.037} & \nError{\pm 1.164}{0.059} \\
                  & \ueeDec           & \nError{-2.527}{0.049} & \nError{\pm  1.180}{0.079} \\
\upsNDec{3}{1};   & \uuuDec           & \nError{-2.426}{0.085} & \nError{\pm  1.313}{0.159} \\
                  & \ueeDec           & \nError{-2.524}{0.093} & \nError{\pm  1.070}{0.153} \\
\hline
\upsCDec{2}{1};   & \uuuDec           & \nError{-0.656}{0.126} & \nError{\pm  0.431}{0.089} \\
                  & \ueeDec           & \nError{-0.689}{0.147} & \nError{\pm  0.425}{0.102} \\
\upsNDec{2}{1};   & \uuuDec           & \nError{-0.148}{0.280} & \nError{    ~~0.000}{1.655} \\
                  & \ueeDec           & \nError{-0.293}{0.330} & \nError{\pm  0.001}{1.130} \\
\hline
\upsNDec{3}{2};   & \uuuDec           & \nError{-0.283}{0.305} & \nError{\pm  0.001}{1.708} \\
                  & \ueeDec           & \nError{-0.583}{0.082} & \nError{\pm  0.003}{1.475} \\
\hline \hline
Simultaneous Fits &                   & \ReBA                  & \ImBA \\
\hline
\multicolumn{2}{l}{ \upsDec{3}{1} } & \nError{-2.523}{0.031} & \nError{\pm  1.189}{0.051} \\
\multicolumn{2}{l}{ \upsDec{2}{1} } & \nError{-0.753}{0.064} & \nError{\pm  0.000}{0.108} \\
\multicolumn{2}{l}{ \upsDec{3}{2} } & \nError{-0.395}{0.295} & \nError{\pm  0.001}{1.053} \\
\hline \hline
\end{tabular}
   }
   \caption{Fit results from $\upsDec{n}{m}$ transitions for 
${\cal B}/{\cal A}$ with ${\cal C}$ set to zero.  The upper
            set of results is from individual fits to each separate
            decay mode and the lower set of results is from
            simultaneous fits to both lepton final states and both
            pion charge modes.  We cannot fit the \upsCDec{3}{2} 
	    transitions,
            individually in $e^{+}e^{-}$ and $\mu^{+}\mu^{-}$ or
	    combined, because of their limited statistics.  In the
            simultaneous fits the relative branching fractions are
            allowed to float.  
            Note that we know the 
	    value of the imaginary part of the ratio only to within a sign.}
   \label{tab:sep_sim_fit}
\end{table*}

\begin{figure*}
   \centerline{
     \hspace*{\fill}
     \resizebox{0.70\textwidth}{!}{\includegraphics{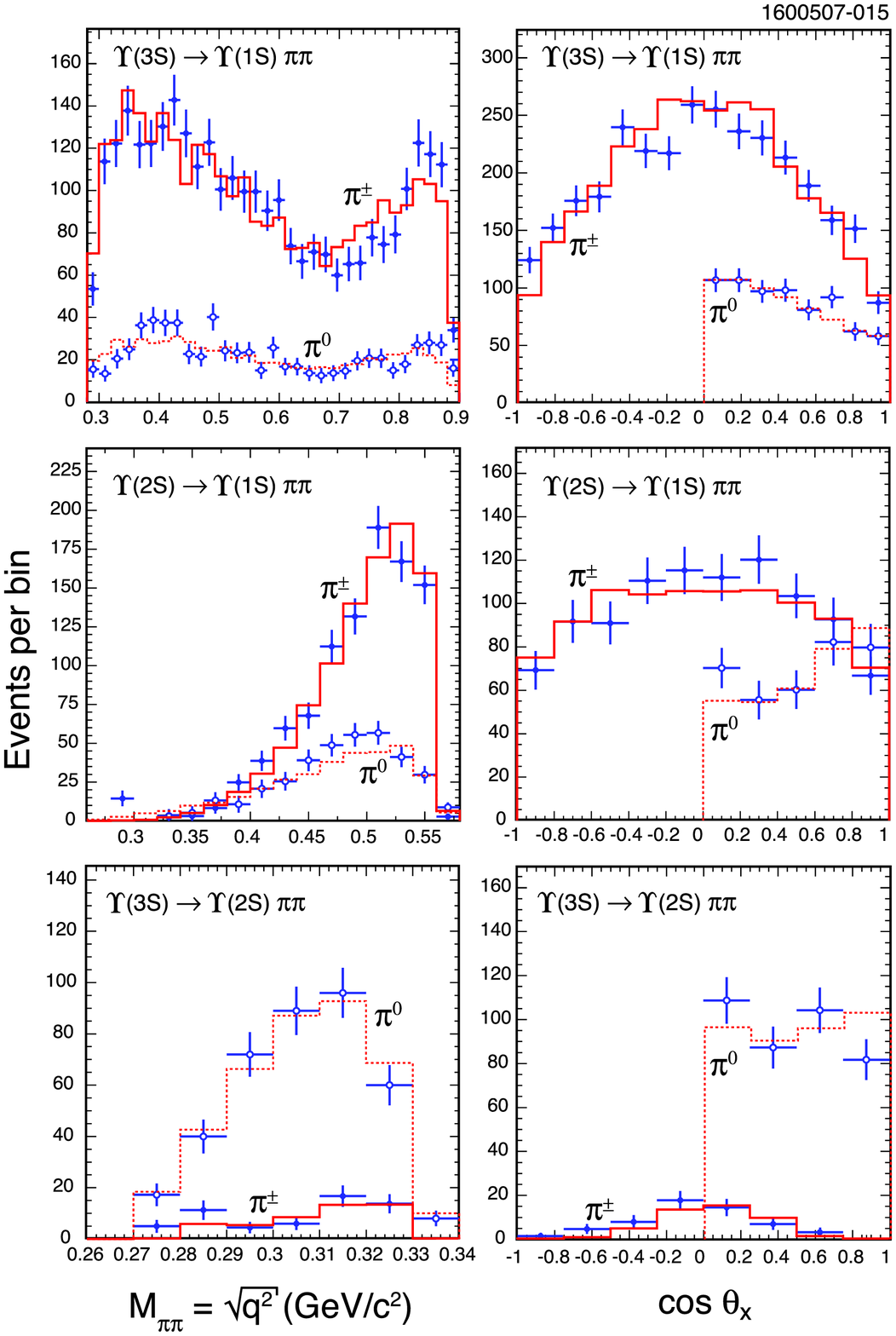}}
     \hspace*{\fill}
   }    
   \caption{ Plots overlaying projections 
	of the data (points with error bars) and the fit result
	(histograms) 
	onto the $\mPiPi$ and
             $\cthx$ variables.  The plots are summed over electrons
             and muons, but are differentiated by pion charge.  The
             neutral modes (open symbols, dashed lines) 
             show only a positive distribution in
             $\cthx$ because the two pions are indistinguishable.  For the 
	     charged modes (solid symbols, solid lines) 
             the angle is that of the $\pi^{+}$.  
               }
   \label{fig:FQFWMC}
\end{figure*}

To study the fit quality 
we 
project the data and the expected decay distribution for the matrix
element value preferred by the fit onto the di-pion mass ($\mPiPi$) and 
di-pion
helicity angle ($\cthx$) 
variables and calculate a $\chi^2$ for each projection.
To increase the bin contents we sum over lepton species but not over
pion charges.  We expect the shapes for charged and neutral pions to
differ due to the rather different 
efficiencies
for reconstruction
and resolutions, 
as
well as the folding of the neutral angle in the fits.  
Figure~\ref{fig:FQFWMC} presents plots of the 
data overlaid with the fit results,
showing good qualitative agreement.
The 
$\chi^2$ values 
from these overlays, given in Table~\ref{tab:chisq},
are
acceptable, given the simplicity of the fitted matrix element.

\begin{table*}
   \centerline{
\begin{tabular}{|c||l|l|l|l|}
\hline
Upsilon & 
\multicolumn{2}{|c|}{$\pi^{+}\pi^{-}$} &
\multicolumn{2}{|c|}{$\pi^{0}\pi^{0}$}\\
\cline{2-5}
Transition & $\cthx$ & $M_{\pi\pi}$& $\cthx$ & $M_{\pi\pi}$ \\
\hline
$3S \to 1S$ & 33.2~(16)& 46.9~(32) & 4.3~(8) & 52.1~(32)\\
$2S \to 1S$ & 6.1~(10) & 22.7~(12) & 3.4~(5) & 13.7~(12)\\
$3S \to 2S$ & 7.1~(7)  & 7.8~(6)   & 7.4~(4) &  2.5~(7) \\
\hline
       \end{tabular}
   }
   \caption{The figure of merit for each of the twelve projections in the 
accompanying figure.  For each projection we give the value of $\chi^{2}$ 
and, in parentheses, the number of bins used to calculate it. 
Uncertainties
in the fit results due to limited simulation statistics are not included
in these calculations.}
   \label{tab:chisq}
\end{table*}

As a further fit quality test, we examine the two dimensional
distribution over the Dalitz variables of error-normalized 
deviations.  
The deviations, 
$\delta_{i}$,
are the difference, 
fit subtracted from the data,
divided
by the mutual uncertainty:
\begin{equation}
   \delta_i = \frac{d_i - \tilde{d}_i}{\sigma_i},
\end{equation}
where 
each
$\tilde{d}_{i}$ 
is the predicted decay population in bin $i$.  The
bin-by-bin uncertainties, $\sigma_i$, are composed of the
uncertainty on the 
data yield in the bin, 
$\sigma_{d} = \sqrt{d_i}$,
and the uncertainty on the template function, dominated by the
fluctuation in the Monte Carlo phase space yield and proportional to
$1/\sqrt{a_i}$, where $a_i$ is the Monte Carlo phase space yield in
bin $i$.  Hence,
   $\sigma_i = \sqrt{d_i + \tilde{d}_{i}^{2}/a_i}$.

The bins for which $d_i = 0$ require special treatment,  
and $\sigma_i$ is modified appropriately.
To minimize the effect of such bins with zero yield, we sum over muon and
electron final states.  This takes a weighted average over the
distributions, rather than taking account of the differences between
the individual distributions and their individual template
predictions.  

\begin{figure*}
   \centerline{
      \hspace*{\fill}
     \resizebox{0.90\textwidth}{!}{\includegraphics{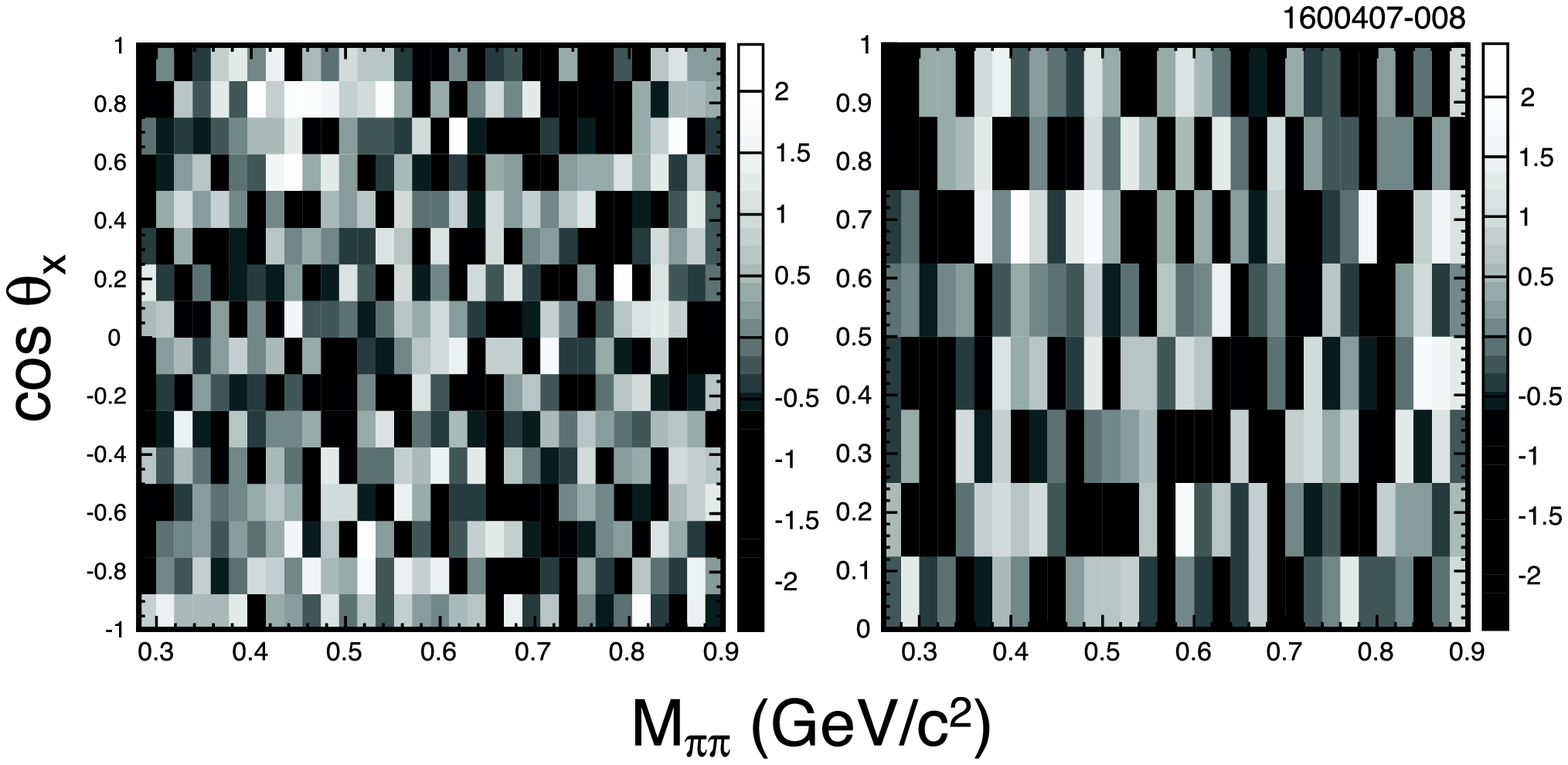}}
      \hspace*{\fill}
   }   

\caption[Error Normalized Fit Deviations over $\cthx$ and $\mPiPi$]
        {Plots of the bin by bin deviations of the data from the fit
        templates normalized to the expected uncertainty on the bin
        content for the transitions between $\Upsilon$(3S) and
        $\Upsilon$(1S).  The left plot is for the charged pion modes while
        the right plot is for the neutral pion modes.  The data are
        summed over lepton species.  No strong concentration of
        deviations is apparent.}
\label{fig:dev_yyy_xx}
\end{figure*}

The deviations between the data and the fit templates, $\delta_{i}$, 
are shown in
Fig.~\ref{fig:dev_yyy_xx} for the charged and neutral transitions
between $\Upsilon$(3S) and $\Upsilon$(1S).  
No significant bunching
is observed that would indicate a bias.  We neglect the small accumulations
in the areas of low tracking efficiency (at large $|\cthx|$ and
intermediate $\mPiPi$), probably attributable to the
Monte Carlo detector model not being 
sufficiently accurate.

\subsection{Fits Including the Chromo-magnetic Term ${\cal C}$}

The fit results in Table~\ref{tab:sep_sim_fit} do not take into
account the possible presence of amplitude terms that come from
chromo-magnetic couplings, which would allow the additional $\clg{C}$
term to appear.  This term is nearly degenerate with the $\clg{B}$
term, and fits allowing it to float show a strong covariance between
these two terms.  This is caused by the similarity in structure of the
two terms; $\clg{B}$ accompanies a functional dependence $E_1 E_2$,
while $(\epsilon'\cdot q_{1,2}) (\epsilon\cdot q_{2,1})$ emphasizes the
regions of phase space in which the pion spatial momentum, and hence
also the energy, are large.  
The low yield modes do not allow the
measurement of the term at all.  We therefore only study it in the
$\upsDec{3}{1}$ transitions, and then only extract a value from the
simultaneous fit.

\begin{figure*}
   \centerline{
      \resizebox{0.6\textwidth}{!}{\includegraphics{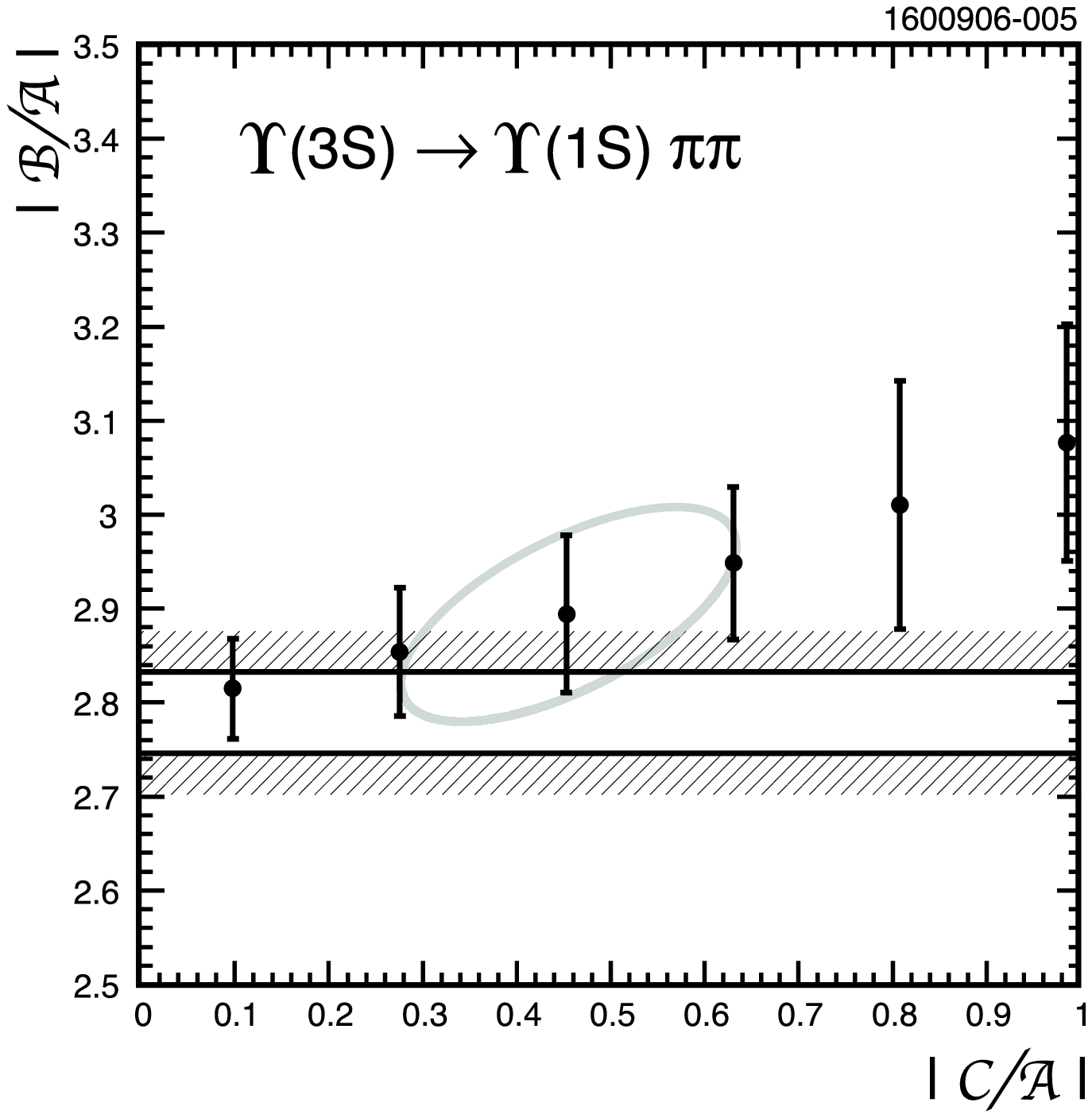}}
      } 
   \caption{Variation of $\clg{B}$ with $\clg{C}$ magnitudes.  The
      points indicate the fit and error for $\clg{B}$ at fixed values
      of $\clg{C}$.  The ellipse indicates the one sigma bound on the
      free fit, the axis of which agrees well with the point by point
      fits.  The bands indicate the one standard error limits on
      $\clg{B}$ when $\clg{C}$ is fixed to zero.  
       }
   \label{fig:bc_fit_scan}
\end{figure*}

The covariance between $\clg{B}$ and
$\clg{C}$ for the \upsDec{3}{1} transition 
is summarized in Fig.~\ref{fig:bc_fit_scan}, which shows
the variation of extracted $|\BoAf|$ with $|\CoAf|$, both as a fit
error ellipse, and as fit trials with $|\clg{C}|$ constrained to
different values.
The ellipse corresponding to one standard deviation from the best
fit gives a
value for 
\upsDec{3}{1} of 
$|\CoAf| = 0.45 \pm 0.18$, with the
uncertainty being purely the statistics of the fit.
The fit which includes real and imaginary
parts of $\CoAf$ shows an improvement over the one with
${\cal C}$ fixed at zero of $-2 \ln \clg{L} = 9.4$. Although this 
implies a $\sim 3\sigma$ improvement in fit quality when ${\cal C}$ is
allowed to float, systematic uncertainties, which are
significant,  have not yet been
taken into account.

With this extended fit the six projections of Fig.~\ref{fig:FQFWMC}
show no significant changes,
and
for the \upsDec{3}{1} transition
the best fit value
of $|\BoAf|$ changes minimally from 2.79 (${\cal C} = 0$)  
to 2.89 (${\cal C}$ floating).
The phase of $\clg{B}$ with respect to $\clg{A}$, denoted 
$\delta_{BA}$, changes little 
(about 2 degrees)
from the 155 degrees of the 
fit done with ${\cal C} = 0$.
The smallness of the effects is not surprising as
the shapes of the $\clg{B}$ and $\clg{C}$ components of the amplitude
are nearly degenerate.
A non-zero value of $|\CoAf|$ may be a consequence of statistical
fluctuations and small systematic biases or may be due to 
$\clg{A}$ and $\clg{B}$ having some dependence on $q^{2}$ and/or
$r^{2}$, {\it i.e.,} not being complex constants.

\subsection{Partial Wave Decomposition}

Since the focus of this study is the decay dynamics of the di-pion
system it is useful to think about the spin structure of the di-pion
composite.  The idea is to look for signatures of higher spin
resonances in the form factors $\clg{A}$ and $\clg{B}$.  We must
account for the intrinsic spin structure of the Lorentz amplitude to
do this.  We equate the Lorentz amplitude with the general partial wave
amplitude to relate the matrix elements.

The transition is of the form $\left<\Upsilon;X|\Upsilon'\right>$.  If
the di-pion system has spin $J$ we have:
\begin{equation}
   \left<1,m_{\Upsilon}; J_X, m_X|1,m_{\Upsilon'}\right>~.
\end{equation}
In that here we assume that only {\cal A} and {\cal B} are non-zero,
there is no change in the polarization from the initial state to
final state Upsilon; more general partial wave decompositions can also be 
made \cite{CKK, newVoloshin}.
The 
angular momentum projections are then
$m_{\Upsilon'} = m_{\Upsilon}$, and $m_X = 0$.
Hence the partial wave decomposition of the $X$ system can only have
$m = 0$ components.  Since the pions are in an iso-singlet state,
their parities require their relative orbital angular momentum to be
even, and hence the orbital angular momentum between the final state
upsilon and the di-pion composite must also be even.  We can only have
even partial waves in our decomposition:
\begin{equation}
   \begin{array}{rcl}
      \clg{M}_P &=& \clg{S}(q^2) Y^0_0 + \clg{D}(q^2) Y^0_2 \\
                &=& \clg{S}(q^2) \frac{1}{\sqrt{4\pi}} +
                    \clg{D}(q^2) \sqrt{\frac{5}{4\pi}}\left(
                         \frac{3}{2} \cos^2 \theta_X - \frac{1}{2}\right).
   \end{array}
\end{equation}
The functions $\clg{S}(q^2)$ and $\clg{D}(q^2)$ are composed of two
terms each, one from the $\clg{A}$ dependence and one from the
$\clg{B}$ dependence:
\begin{equation}
   \clg{S}(q^2) = \clg{A}\clg{S}_{\clg{A}}(q^2) +
                  \clg{B}\clg{S}_{\clg{B}}(q^2),
   \hspace*{5mm}{\rm and}\hspace*{5mm}
   \clg{D}(q^2) = \clg{A}\clg{D}_{\clg{A}}(q^2) +
                  \clg{B}\clg{D}_{\clg{B}}(q^2).
\end{equation}
We here assume that there are no significant contributions from
partial waves higher
than $J = 2$.  This will be true if there are no contributions from
variations of form factors over the Dalitz space.  Higher $J$ terms
must originate from structure in the form factors $\clg{A}$ and
$\clg{B}$.

Equating the decay distributions (or equivalently, projecting inner
products over the angular space) yields the following forms:
\begin{equation}
      {\clg{S}}_{\clg{A}}(q^2) = q^2 - 2 M_\pi^2~,
   \hspace*{7mm}{\rm and}\hspace*{7mm}
      {\clg{D}}_{\clg{A}}(q^2) = 0
\end{equation}
for a pure ``$\clg{A}$'' decay, and
\begin{widetext}
\begin{equation}
   \begin{array}{rcl}
      {\clg{S}}_{\clg{B}}(q^2) & = &
                 \frac{q^2 \left( {\left( M_{\Upsilon'}^2 -
                 M_{\Upsilon}^2 \right) }^2 +
                 \left( M_{\Upsilon'}^2 + M_{\Upsilon}^2 \right)
                 q^2 - 2 {q^2}^2 \right)  +
                 2 M_{\pi}^2 \left( {M_{\Upsilon'}^2}^2 +
                 {\left( M_{\Upsilon}^2 - q^2 \right) }^2 -
                 2 M_{\Upsilon'}^2 \left( M_{\Upsilon}^2 + q^2
                 \right)  \right) }{12
                 {\sqrt{M_{\Upsilon'}^2 M_{\Upsilon}^2}} q^2}~; \\
      {\clg{D}}_{\clg{B}}(q^2) & = &
                 \frac{\left( 4 M_{\pi}^2 - q^2 \right)
                 \left( {M_{\Upsilon'}^2}^2 +
                 {\left( M_{\Upsilon}^2 - q^2 \right) }^2 -
                 2 M_{\Upsilon'}^2 \left( M_{\Upsilon}^2 + q^2 \right)
                 \right) }{12 {\sqrt{5}}
                 {\sqrt{M_{\Upsilon'}^2 M_{\Upsilon}^2}} q^2}
   \end{array}
\end{equation}
\end{widetext}
for a pure ``$\clg{B}$''~decay.  The overall amplitude is
\begin{equation}
   \clg{M}_P = ( \clg{A}\,{\clg{S}}_{\clg{A}}(q^2) +
                 \clg{B}\,{\clg{S}}_{\clg{B}}(q^2)   ) Y^0_0 +
               ( \clg{A}\,{\clg{D}}_{\clg{A}}(q^2) +
                 \clg{B}\,{\clg{D}}_{\clg{B}}(q^2)   ) Y^0_2~,
\end{equation}
where it is implied that $Y^m_l$ is a function of the helicity angles
of the pseudo-decay $X \rightarrow \pi \pi$, $\theta_X$ and $\phi_X$
(although the latter variable plays no role in the description of this
decay, by the assumptions above).  Interference between the $S$-wave
and $D$-wave components of the decay comes from the functions
$\clg{S}(q^2)$ and $\clg{D}(q^2)$ being complex valued.  Though
$\clg{S}_{\clg{A},\clg{B}}(q^2)$ and $\clg{D}_{\clg{A},\clg{B}}(q^2)$
are real functions, $\clg{A}$ and $\clg{B}$ are complex coefficients
with nontrivial relative phase.

\begin{figure*}
   \centerline{
      \resizebox{0.8\textwidth}{!}{
         \includegraphics{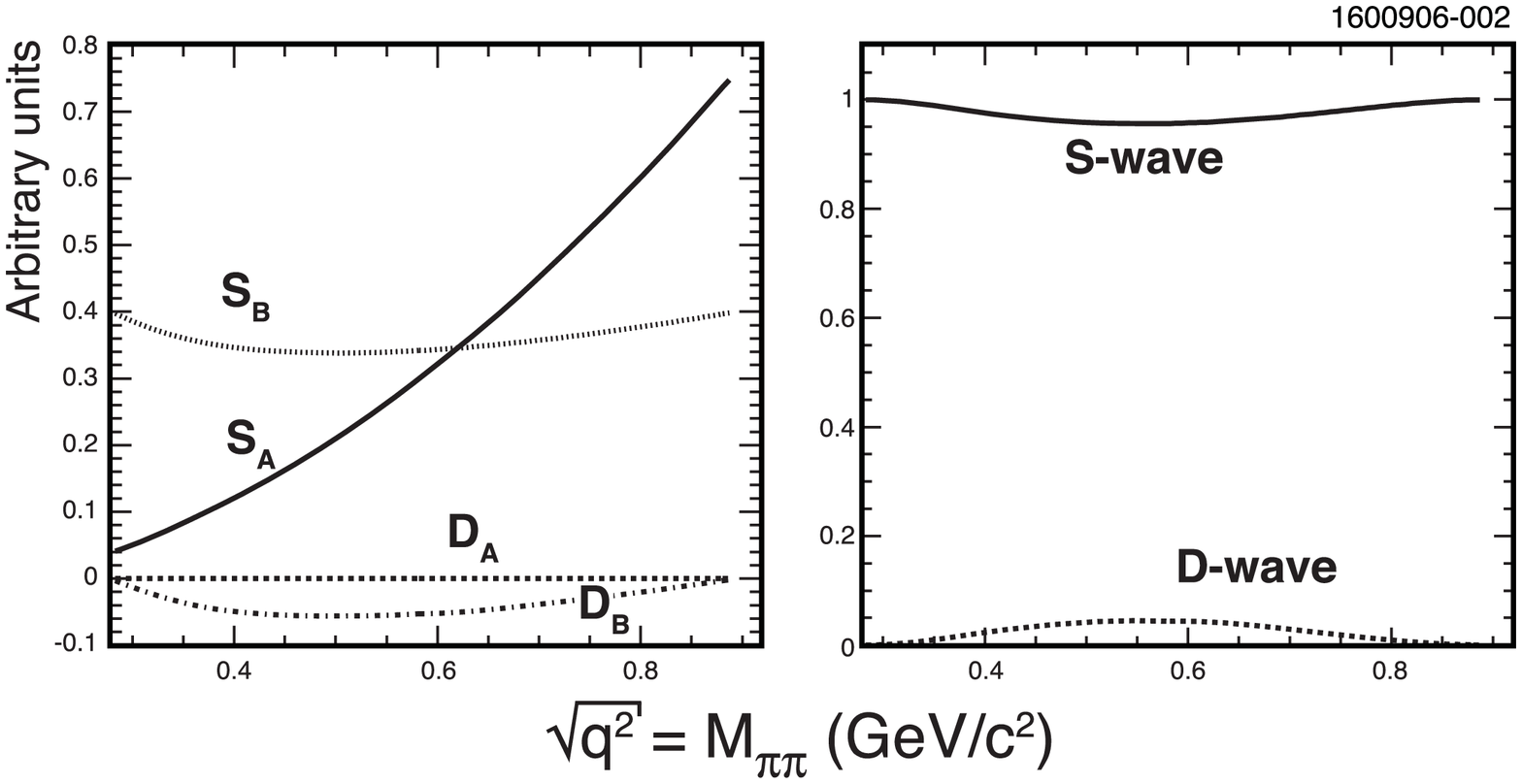}}
   }
\caption[Partial Wave Components] {The left plot shows the amplitude
        component functions ${\clg{S}}_{\clg{A}}$,
        ${\clg{S}}_{\clg{B}}$, ${\clg{D}}_{\clg{A}}$, and
        ${\clg{D}}_{\clg{B}}$ as a function of $\mPiPi \equiv
        \sqrt{q^2}$.  These are summed to obtain the total amplitude.
        The partial rate to $S$-wave and $D$-wave components are shown
        in the right plot for the \upsDec{3}{1} decay as determined
        from the results of this analysis: $\BoAf = -2.52 + 1.19
        i$.  Note that the $D$-wave contribution is largest in the low
        to intermediate range of $q^2$, and is suppressed at both
        extrema by angular momentum barrier effects.  
	Note further
	that this is not a resonance phenomenon
        despite its shape in $\mPiPi$ and the changing angular
        structure.}
\label{fig:pwsVsMpp}
\end{figure*}

The structure of $S$ and $D$ components as functions of $q^2$ are
determined by the assumptions underlying the derivation of the general
Lorentz amplitude.  The four functions from the pure~$\clg{A}$ and
pure~$\clg{B}$ components are sketched in Fig.~\ref{fig:pwsVsMpp}
together with the fractional $S$- and $D$-wave components in the
angular distribution (which can alternately be thought of as the
strengths of the $S$- and $D$-wave components), extracted from our fit
to $\upsDec{3}{1}$.

This partial wave extraction becomes much more complex if the form
factors are assumed to be variable over the Dalitz space, for example
due to resonant structure/enhancement in the decay.  This will
introduce higher powers of $\cos^2 \theta_X$ to the overall amplitude
and will need higher partial wave components to account for the
variation.

The presence of $D$-wave components in the angular
distribution of the decay is not in itself an indication of resonances
contributing, nor the presence of unaccounted-for physics.
The presence of 
a $q^{2}$-dependent
$D$-wave
component could simply be a consequence of angular momentum
barriers in the three body phase space of the decay.  
The data do not demand the introduction of a 
$q^{2}$-dependent magnitude or phase for ${\cal B}/{\cal B}$.
These small $D$-wave components are consistent with those derived in
a recent paper by Voloshin~\cite{Voloshin:2006}, 
in which he emphasizes the importance
of relativistic and chromo-magnetic effects.

\section{Systematic Uncertainties}

We address three sources of systematic uncertainty in the measurements
of $\BoAf$ and $\CoAf$: 
model dependence, detector efficiency and resolution, and
backgrounds.  

In Sect. III we showed that our model provides a very good
description of the data in the $(q^{2},\cthx)$ plane and
that the 
presence or absence of the chromo-magnetic coupled
term in the amplitude has little effect on $|\BoAf|$ and
$\delta_{BA}$.

Uncertainty in the estimation of the detector efficiency and
resolution contributes most significantly in the charged mode analyses
due to our limited knowledge of the tracking efficiency at very low
momentum.  In that the low momentum region is precisely where the matrix
element has potential suppression in the $\clg{B}$ term, this can
potentially cause a significant bias.  
To estimate this effect we use the full Monte Carlo simulation with
looser and tighter track reconstruction requirements to provide bounds
on the shape of the efficiency as a function of track curvature. We then
create a number of analytic functions that span these boundaries.  Then
we use a toy Monte Carlo to simulate events with one of these
analytic functions and assume a different one for the reconstruction.  
The variations in the fit results are conservatively
assumed to be one standard error uncertainties on the extracted
parameters.

The same process is repeated for the neutral modes, varying the
thresholds at which showers can be observed in the detector.  This
obviously leads to a large variation in branching ratios from simple
inability to reconstruct the decays, but does not exhibit any
significant change in the shape of the efficiency function over the
measurement variables.  This is to be expected since the $\pi^0$ decays
have largely flat acceptance over the kinematic range of these decay
modes.

We have evaluated the systematic errors associated with detector
resolution, and find them to be negligible in comparison with the
statistical errors from the fit and the other systematic errors
discussed here.  The curvatures of the matrix element components 
across the Dalitz plot
are all very much smaller than the variances of the reconstructed
measurement variables around their true values.  No systematic
uncertainty is assigned to this source.

Background subtraction is only a source of bias if the upper and lower
sidebands in the recoil mass exhibit markedly different shapes or the
background is strongly peaked under the signal.  In this case the
extrapolations of the background shape and magnitude under the peak
could be distorted.  
We have redone the fits with the ratio of 
the widths of sideband window to signal window
both doubled and halved, and with only using either the high-mass or low-mass
sideband.
The variations in the fit
are conservatively taken to represent one sigma
variations in the final result, and are given in the last 
column of Table \ref{tab:full_result}.

Finally, the lepton reconstruction is capable of contributing bias
since all decay modes are fully reconstructed.  However, the detector
response to leptons is sufficiently well measured in other analyses
that the detector simulation is much more precise than what is
required for this data set.  The variation of the shapes is furthermore
only relevant for the final $\clg{C}$ term, which is dependent on the
lepton polar angle.  With the exception of a small part of the
$\clg{C}$ terms there can be no effect due to lepton acceptance.  We
estimate any systematic error associated with the lepton
reconstruction to be negligible.

The fit results combined with these systematic uncertainties are
summarized in Tables~\ref{tab:full_result}
and~\ref{tab:full_result_summary}.  Since the magnitude $|\CoAf|$ in
the fit is only separated from zero by about one standard error and is
expected to be suppressed in the theoretical models, we set a limit
rather than claim observation of a non-zero value.

We set this limit by assuming the value of $\CoAf$ has a Gaussian
uncertainty in real and imaginary parts.  We transform variables to
$|\CoAf|$ and $\arg(\CoAf)$, using the sum of the variances of
statistical and systematic origin as the overall variance.  
We then find the 90\% 
upper 
limit
from the resulting
distribution as
\begin{equation}
   \begin{array}{rclcll}
      |\CoAf| & < & 1.09 & \hspace*{1cm} & \rm{at}~90\%~C.L.~.  &
   \end{array}
\end{equation}

\section{Summary and Acknowledgments}

We quote fit results for the three transitions from simultaneous fits
to the different decay modes with statistical and systematic
uncertainties in Table~\ref{tab:full_result_summary}.  Only the simplest
features of the Brown and Cahn decay amplitude
(Eqn.~\ref{Eq:BandCAmplitude}) are included in our model, and the
fits account for the structure of the decay without introduction of
new physics or 
contributions from
resonances. 

\begin{figure*}
   \centerline{ 
     \resizebox{0.5\textwidth}{!}{\includegraphics{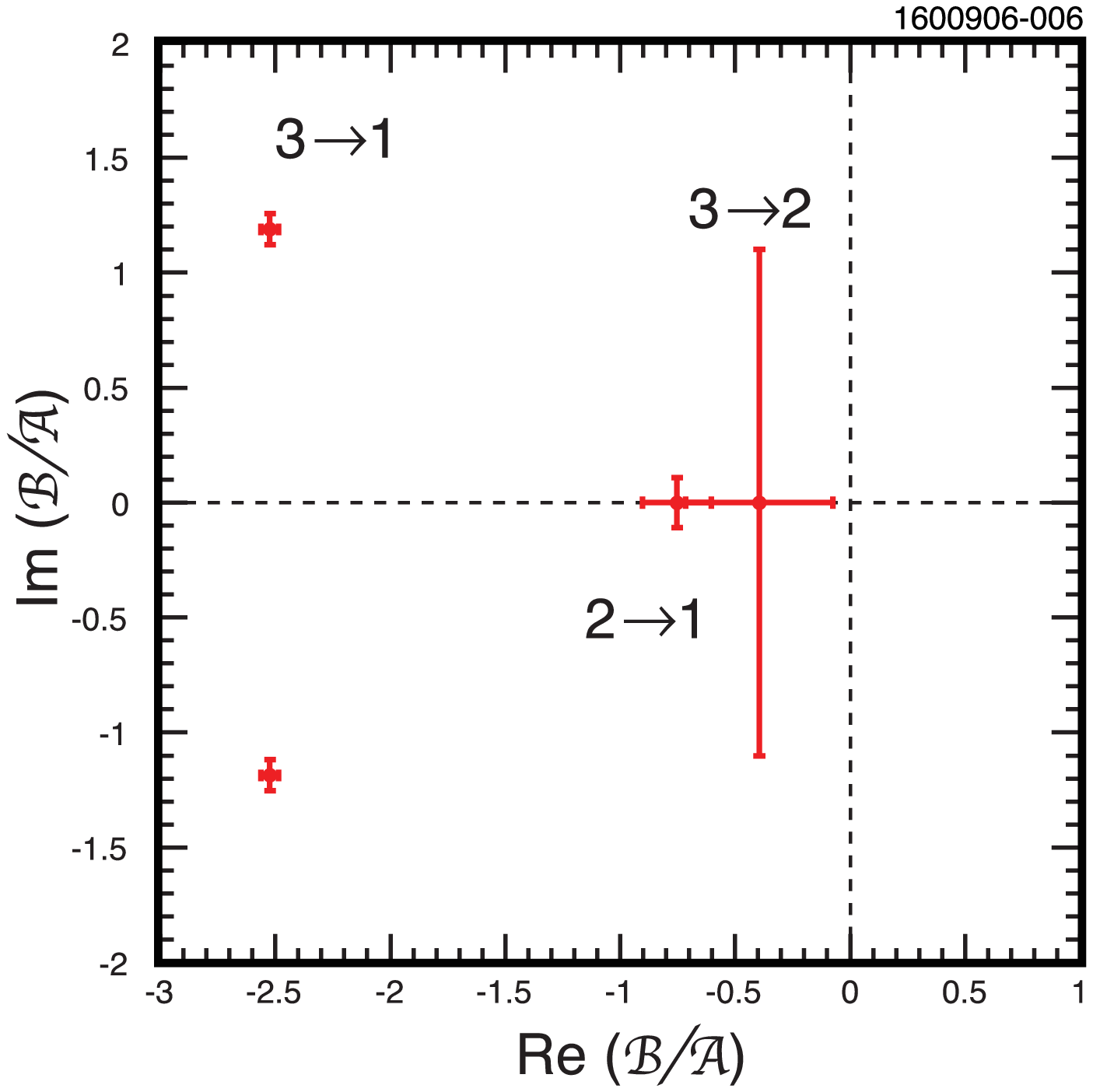}}
   } 
   \caption{ 
Complex values of matrix element ratio
             $\clg{B}/\clg{A}$ from combined fits for the three
             transitions under the assumption that ${\cal C} = 0$.  Note the
             two-fold ambiguity in the imaginary part.}
   \label{fig:matels_vs_plots}
\end{figure*}

The matrix elements are indicated as points in the complex plane in
Fig.~\ref{fig:matels_vs_plots}.  For the ``anomalous''
$\upsDec{3}{1}$ transition we fit for the presence of the ``suppressed''
$\clg{C}$ term as a test for the breakdown of the underlying
assumptions leading to the standard matrix element.  
This term is not significant when systematic errors are taken into account and
the quality of the fit to the data is good without it. 
Therefore, we set an upper limit of
$|\CoAf| < 1.09$ at $90\%$ C.L..  

We note in particular that the treatment of the di-pion transitions
via the full allowed matrix element under the assumptions 
in Refs.~ 
\cite{Brown:1975dz,MBVoloshin:75,Gottfried:1977gp,Yan:1980uh,Voloshin:1980zf}
allows two matrix elements, only one of which has
traditionally been assumed to be non-zero.  The description of the
$\upsDec{3}{1}$ transition di-pion mass and angular structure as
anomalous is only true in the limit of this assumption.  
This analysis
shows in particular that the description of the decay process in terms
of the two favored amplitude terms, 
with 
complex
form factors constant over the Dalitz plane,
suffices to describe the decay distributions of
$\upsDec{3}{1}$, $\upsDec{3}{2}$, and $\upsDec{2}{1}$,
provided the form factors are allowed to vary with the transition.
For the $\Upsilon$(3S)$\to\Upsilon$(1S)$\pi\pi$ transition,
we find $|{\cal B}/{\cal A}|= 2.79 \pm 0.05$, which could imply
a large magnitude of ${\cal B}$ or a suppressed ${\cal A}$; recent
theoretical considerations~\cite{Voloshin:2006} 
favor the latter interpretation.
While smaller than in the case of $\upsDec{3}{1}$, $|\BoAf|$ is
also determined to be non-zero for the  case of $\upsDec{2}{1}$.
The large imaginary part of ${\cal B}/{\cal A}$ is 
intriguing~\cite{newVoloshin}.

While there are not yet first principles predictions of the values of
the matrix elements of the decays studied here, this analysis does
provide complete measurements of the relative matrix element
magnitudes and phases that can serve as a point of comparison with
{\it ab initio} QCD calculations.

\begin{table*}
   \centerline{
\begin{tabular}{llccccc}
\hline \hline
Fit, No $\clg{C}$   &       &        & stat.       & effcy.  ($\pi^\pm$)   & effcy.($\pi^0$) & bg.  sub.    \\
\hline
\upsDec{3}{1}
   & $\begin{array}{c} \ReBA \\ \ImBA \end{array}$
   & $\begin{array}{r} -2.523 \\  \pm 1.189 \end{array}$
   & $\begin{array}{l} \pm 0.031 \\ \pm 0.051 \end{array}$
   & $\begin{array}{l} \pm 0.019 \\ \pm 0.026 \end{array}$
   & $\begin{array}{l} \pm 0.011 \\ \pm 0.018 \end{array}$
   & $\begin{array}{l} \pm 0.001 \\ \pm 0.015 \end{array}$ \\
\hline
\upsDec{2}{1}
   & $\begin{array}{c} \ReBA \\ \ImBA \end{array}$
   & $\begin{array}{r} -0.753 \\ 0.000 \end{array}$
   & $\begin{array}{l} \pm 0.064 \\ \pm 0.108 \end{array}$
   & $\begin{array}{l} \pm 0.059 \\ \pm 0.036 \end{array}$
   & $\begin{array}{l} \pm 0.035 \\ \pm 0.012 \end{array}$
   & $\begin{array}{l} \pm 0.112 \\ \pm 0.001 \end{array}$ \\
\hline
\upsDec{3}{2}
   & $\begin{array}{c} \ReBA \\ \ImBA \end{array}$
   & $\begin{array}{r} -0.395 \\ \pm 0.001 \end{array}$
   & $\begin{array}{l} \pm 0.295 \\ \pm 1.053 \end{array}$
   & 
   & $\begin{array}{r} \pm 0.025 \\ \pm 0.180 \end{array}$
   & $\begin{array}{l} \pm 0.120 \\ \pm 0.001 \end{array}$ \\
\hline \hline
Fit, float $\clg{C}$   &       &        & stat.       & effcy.  ($\pi^\pm$)   & effcy.($\pi^0$) & bg.  sub.    \\
\hline
\upsDec{3}{1}
   & $\begin{array}{c} |\BoAf| \\ |\CoAf| \end{array}$
   & $\begin{array}{r} 2.89 \\ 0.45 \end{array}$
   & $\begin{array}{l} \pm 0.11 \\ \pm 0.18 \end{array}$
   & $\begin{array}{l} \pm 0.19 \\ \pm 0.28 \end{array}$
   & $\begin{array}{l} \pm 0.11 \\ \pm 0.20 \end{array}$
   & $\begin{array}{l} \pm 0.027 \\ \pm 0.093 \end{array}$ \\
\hline \hline
\end{tabular}
   }
   \caption{Combined fit results for all transitions with statistical and
   systematic uncertainties.  The systematic uncertainties are in
   order: 
   $\pi^\pm$ detection efficiency, 
   $\pi^0$ detection efficiency, and
   background subtraction for the \upsDec{3}{2} transition.
   The upper set of 
   results are for the fits assuming contributions to the amplitude
   from only the $\clg{A}$ and $\clg{B}$ terms.  The bottom two lines
   are the fit results when the $\clg{C}$ term is allowed to be
   non-zero.  The imaginary part of the ratio has a two-fold ambiguity and
	is only known to within a 
	sign.  Note that for the transition $\upsDec{3}{2}$ we do
   not have fits for the charged di-pion case.}
   \label{tab:full_result}
\end{table*}

\begin{table*}
   \centerline{
\begin{tabular}{llc}
\hline \hline
Fit, no $\clg{C}$, total error & & \\
\hline
\upsDec{3}{1}
   & $\begin{array}{c} \ReBA \\ 
                       \ImBA \\ 
                       |\BoAf| \\ 
                       \delta_{BA} \end{array}$
   & $\begin{array}{rcl} -2.52 & \pm & 0.04 \\ 
                          \pm 1.19 & \pm & 0.06 \\
                          2.79 & \pm & 0.05 \\
                          155(205)  & \pm &  2  \end{array}$ \\
\hline
\upsDec{2}{1}
   & $\begin{array}{c} \ReBA \\
                       \ImBA \\
                       |\BoAf| \\ 
                       \delta_{BA} \end{array}$
   & $\begin{array}{rcl} -0.75 & \pm & 0.15 \\
                         0.00 & \pm & 0.11 \\
                          0.75 & \pm & 0.15 \\
                          180  & \pm &  9  \end{array}$ \\
\hline
\upsDec{3}{2}
   & $\begin{array}{c} \ReBA \\ \ImBA \end{array}$
   & $\begin{array}{rcl} -0.40 & \pm & 0.32 \\ 0.00 & \pm & 1.1 \end{array}$ \\
\hline \hline
Fit, float $\clg{C}$, total error & & \\
\hline
\upsDec{3}{1}
   & $\begin{array}{c} |\BoAf| \\ |\CoAf| \end{array}$
   & $\begin{array}{rcl}  2.89 & \pm & 0.25 \\  0.45 & \pm & 0.40 \end{array}$ \\
\hline \hline
\end{tabular}
   }
\caption{Fit results for all transitions with total uncertainties.
  These numbers represent the final result of this analysis.  In the 
case of the magnitude ratio $|\CoAf|$, we also quote a limit as detailed
in the text.  The phase angles are quoted in degrees, and have a two-fold
ambiguity of reflection in the real axis.}
\label{tab:full_result_summary}
\end{table*}

We gratefully acknowledge the effort of the CESR staff
in providing us with excellent luminosity and running conditions.
D.~Cronin-Hennessy and A.~Ryd thank the A.P.~Sloan Foundation.
This work was supported by the National Science Foundation,
the U.S. Department of Energy, and
the Natural Sciences and Engineering Research Council of Canada.

\newpage

\section{Appendix: Details of the Likelihood Fitter}

This appendix gives some details of our application of the
likelihood fitter.

Smearing due to reconstruction resolution adds a small variance to the
Poisson error on the Monte Carlo integral, but the smearing widths are
small compared to the scales over which the matrix element changes so
this additional variance is small.  For any shape with an
approximately polynomial form at a point, the resolution is described
by convolving a Gaussian with the polynomial.  As an example, we
assume a functional form $g^T = a + b\,x + c\,x^2$ and seek its
observed shape in terms of the observed variables, $g^O( x^O )$,
using a Gaussian transformation:
\begin{eqnarray}
   g^O( x^O ) & = & \int dx^T \clg{G}( x^T - x^O | \mu \equiv 0, \sigma )
                    g^T( x^T ) \\
              & = & \int dx^T \clg{G}( x^T - x^O | \mu \equiv 0, \sigma )
                   ( a + b\,x^T + c\,( x^T )^2 ) \\
              & = & ( a + c\,\sigma^2 ) + b\,x^O + c\,(x^O)^2
\end{eqnarray}
So long as 
$\sigma^2 \ll a/c$, 
\idest the resolution is small compared
to the curvature, the shape will not be materially changed.  For the
angular dependence, which is quartic in $\cthx$ this means the
resolution need only be small compared to 1/2; the observed
resolutions are of the order of 5\% or less.  In $\mPiPi$ the same
holds true, with the scale being given by the pion mass,
\nUnits{140}{\mMeV}, and the observed resolutions being at worst
\nUnits{10}{\mMeV}.  The shape of the decay amplitude is not changed
significantly by these resolutions, but any residual effect is
included in the estimated tracking and shower systematic
uncertainties.

Our problem differs from that discussed in
Ref.~\cite{Barlow:dm} in that the templates do not have independent
Poisson fluctuations.  The underlying phase space simulation has a
Poisson fluctuation, but the templates are known (very nearly) exactly
and uncertainties on them do not contribute to the overall likelihood
function.

In the absence of background this problem is solved as follows,
with each two-dimensional ($q^{2}$, cos$\theta_{X}$) bin denoted by subscript
$i$.  

We compare the Monte Carlo simulated, acceptance and efficiency-corrected,
phase space distribution (with true and observed yields $A_i$
and $a_i$), 
multiplied by the modulus squared of the amplitude, with the data
distribution (with true and observed yields $D_i$
and $d_i$).  
Both distributions are subject to Poisson
fluctuation:
\begin{equation}
    {\cal P}( d_i; D_i ) = \frac{e^{-D_i}{D_i}^{d_i}}{d_i!}
    \hspace*{5mm}\mbox{and}\hspace*{5mm}
    {\cal P}( a_i; A_i ) = \frac{e^{-A_i}{A_i}^{a_i}}{a_i!}.
\end{equation}
Bin-by-bin, the modulus squared of the decay amplitude appears in the exact
relation between the 
true data
yields $D_i$ and the true phase
space yields $A_i$:
\begin{equation}
    D_i = f_i( \alpha )\,A_i.
\end{equation}
The function $f_i$ represents the decay distribution ($|{\cal M}|^{2}$) in the kinematic
space bin $i$ as a function of $\alpha$, the decay parameters.  In
this case $\alpha$ consists of real and imaginary parts of $\BoAf$ and
$\CoAf$.

The log likelihood used in this fit is then given by, summing over all the bins:
\begin{equation}
      \ln {\cal L}(\alpha) = \sum_{i=1}^n \left(
                      d_i \ln{f_i(\alpha) A_i} - f_i(\alpha) A_i - \ln{d_i!}
                       + a_i \ln{A_i} - A_i - \ln{a_i!} \right).
\end{equation}

The $A_i$ represent the phase space subject to efficiency and
acceptance effects and are uninteresting nuisance parameters
that can be eliminated by extremizing the likelihood with respect
to them.  Proceeding
in analogy with the approach in \cite{Barlow:dm} we can find the
analytic extremum condition, solve for $A_i$
\begin{equation}
    A_i = \frac{d_i + a_i}{f_i + 1}
\end{equation}
and substitute back into the likelihood function to give a reduced
likelihood:
\begin{equation}
    \ln {\cal L}(\alpha) = \sum_{i=1}^n \left[ \rule{0pt}{16pt} d_i \ln f_i(\alpha)
                - ( d_i+a_i ) \ln( 1 + f_i(\alpha) ) \right]+ {\rm const.}
\end{equation}
We then minimize $-2 \ln {\cal L}$ with respect to the fit parameters
$\alpha$ (occurring only in the coefficients $f_i$).  This is
implemented using the CERN Library minimization package,
MINUIT~\cite{CERNLib:MINUIT}.

The full likelihood as used in the fit includes an extension of this
approach to account for background under the signal peaks.  This
introduces additional parameters $B_i$ and $b_i$.  These represent bin
by bin true and observed background yields.  The $B_i$ are a second
set of nuisance parameters that are eliminated in the same way as were
the $A_i$ before.  The resulting likelihood is significantly more
complicated in detail but not in principle.  For brevity it is not
included here.

\newpage

%
%

\end{document}